\newcommand{\sonec}{1.05} %size figure one column
\newcommand{\stwoc}{1.00} %size figure two column

\def\figa{
  \begin{figure}[!t] 
  \centering
  \includegraphics[scale=\sonec]{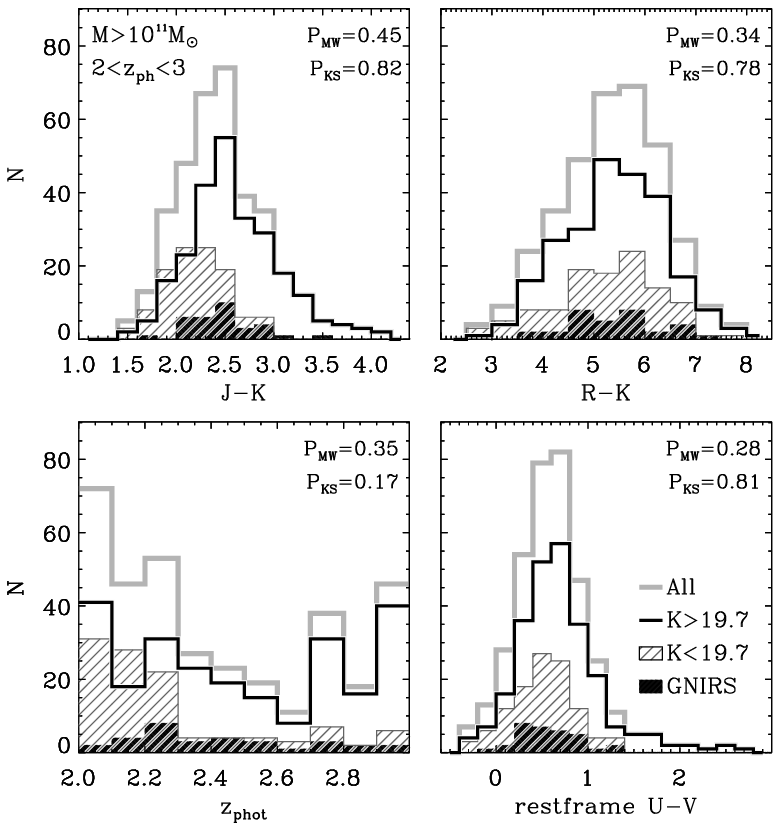} 
  \figcaption{Comparison between the photometric properties of the
    GNIRS sample at $2<z_{\rm phot}<3$ and a mass-limited sample
    ($M>10^{11} M_{\odot}$) at $2<z_{\rm phot}<3$.  The probabilities
    ($P$) that the GNIRS sample and the full mass-selected sample have
    similar distributions, as derived using a Mann-Whitney (MW) and a
    Kolmorov-Smirnov (KS) test, are given in the panels. Additionally,
    we divide the mass-selected sample into its $K$-bright ($K<19.7$)
    and $K$-faint ($K>19.7$) members. The GNIRS sample may be less
    representative for a $K$-bright sample, as the redshift
    distribution is different. \label{mass_vs_k_sel} }
  \end{figure} 
}

\def\figb{
  \begin{figure}[!t]
  \centering
  \includegraphics[scale=\sonec]{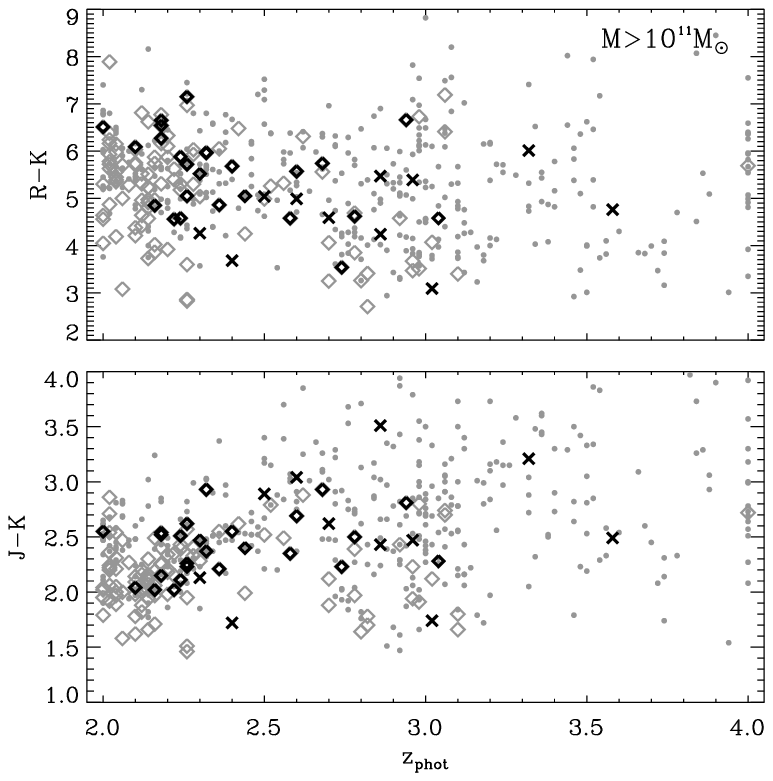} 
  \figcaption{Comparison of $J-K$ and $R-K$ colors as a function of
    \zp\ between a photometric mass-limited ($M>10^{11} M_{\odot}$)
    sample and our spectroscopic sample. The gray diamonds and dots
    represent all massive galaxies in the deep MUSYC fields (SDSS1030,
    1256 and HDF-S) with $K<19.7$ and $K>19.7$ respectively. The black
    symbols represent the 36 galaxies of the GNIRS sample, selected
    from both the MUSYC deep ({\it diamonds}) and wide (ECDFS, {\it
    crosses}) surveys. \label{rep} }
  \end{figure}
}

\def\figc{
  \begin{figure}[!t]
  \centering
  \includegraphics[scale=\sonec]{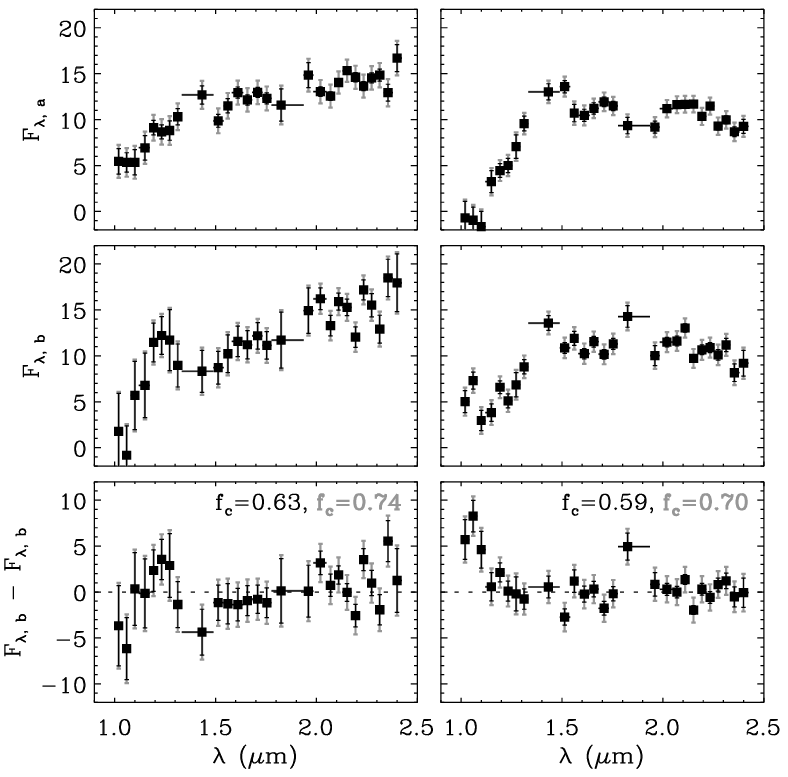}
  \figcaption{The differences between the low resolution spectra of
    two observing sequences for the same galaxy ({\it left:} 1030-301,
    {\it right:} 1256-0) are used to estimate a systematic uncertainty
    on the spectra. In order to improve the consistency between the
    different observing sequences, we increase the original
    uncertainty per bin ({\it black error bars}) by a systematic error
    of 10\% ({\it gray error bars}) of the average flux in the binned
    spectrum. The fractions of bins that are consistent within
    1$\sigma$ for the original and increased uncertainties are given
    in the panels in black and gray respectively. \label{error}}
  \end{figure}
}

\def\figd{
  \begin{figure}[!t]
  \centering
  \includegraphics[scale=\sonec]{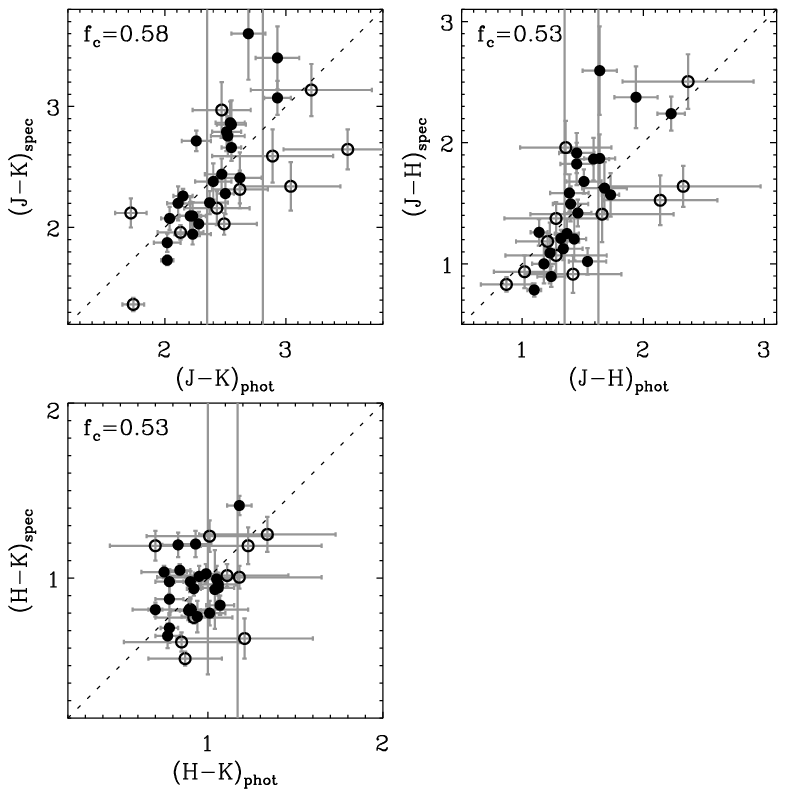} 
  \figcaption{Comparison of photometric NIR colors, and those derived
    from the spectra, for galaxies in the deep ({\em solid}) and wide
    ({\em open}) MUSYC fields. Both colors are not corrected for flux
    contributions by emission lines. For a few galaxies the spectra
    are too noisy to measure the spectroscopic NIR colors. The
    fraction of galaxies for which the photometric and spectroscopic
    colors are consistent within 1$\sigma$ is given in the top left
    corner. As these factors are less than $0.68$, the errors may be
    slightly underestimated.  \label{q_check}}
  \end{figure} 
}

\def\fige{
  \begin{figure*}[!t] \centering
  \includegraphics[scale=1]{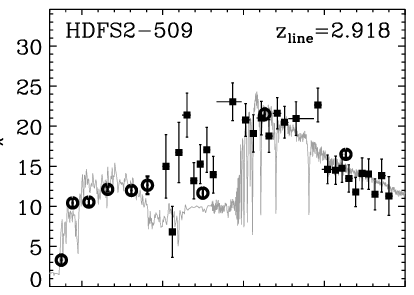}
  \includegraphics[scale=1]{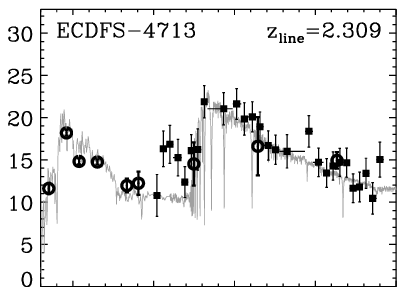}
  \includegraphics[scale=1]{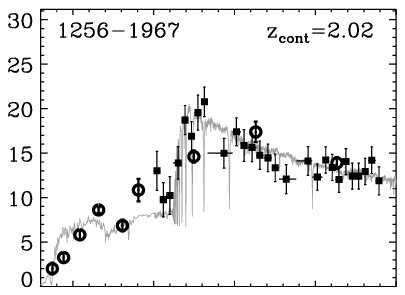}
  \includegraphics[scale=1]{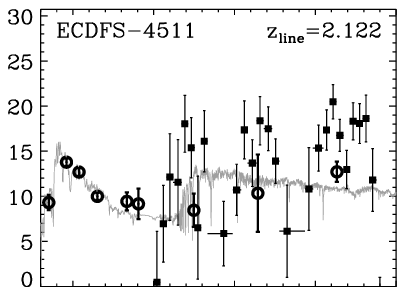}\\
  \vspace{-0.02in} \includegraphics[scale=1]{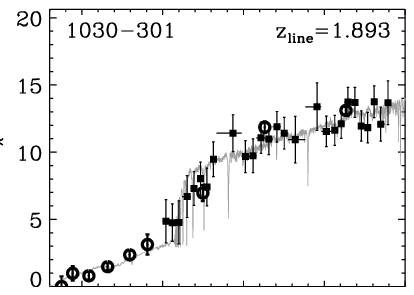}
  \includegraphics[scale=1]{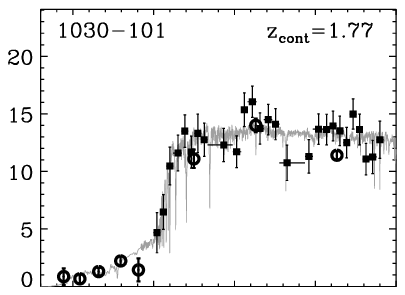}
  \includegraphics[scale=1]{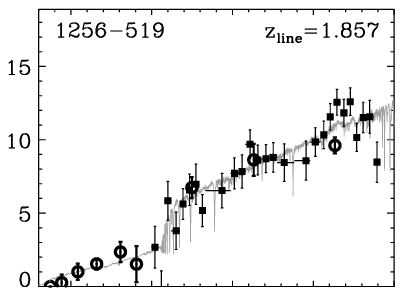}
  \includegraphics[scale=1]{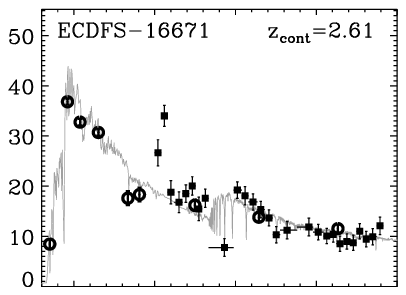}\\
  \vspace{-0.02in} \includegraphics[scale=1]{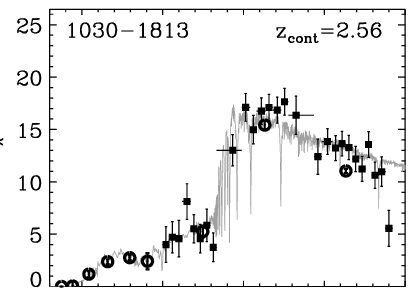}
  \includegraphics[scale=1]{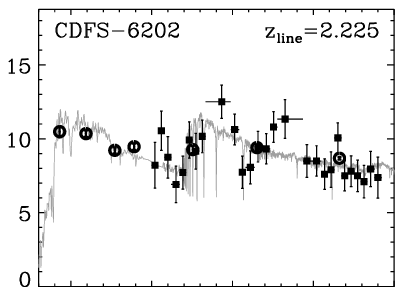}
  \includegraphics[scale=1]{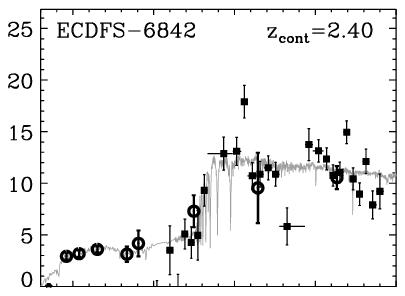}
  \includegraphics[scale=1]{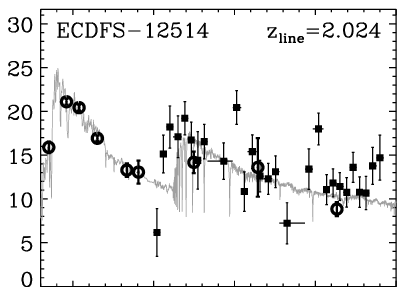}\\
  \vspace{-0.02in} \includegraphics[scale=1]{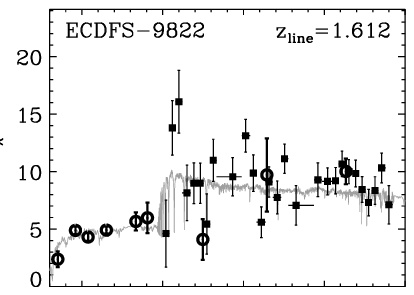}
  \includegraphics[scale=1]{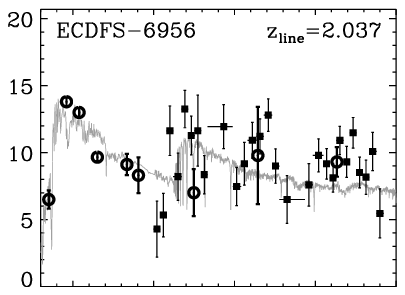}
  \includegraphics[scale=1]{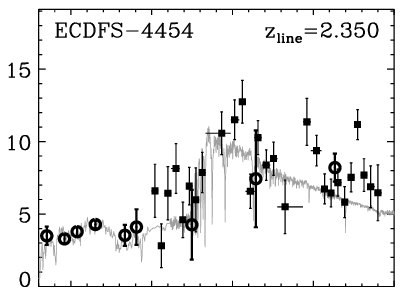}
  \includegraphics[scale=1]{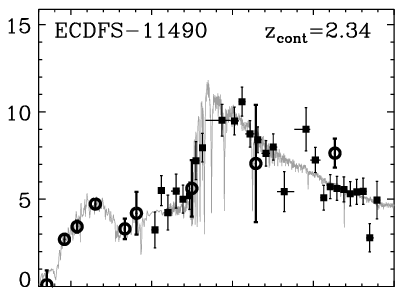}\\
  \vspace{-0.02in} \includegraphics[scale=1]{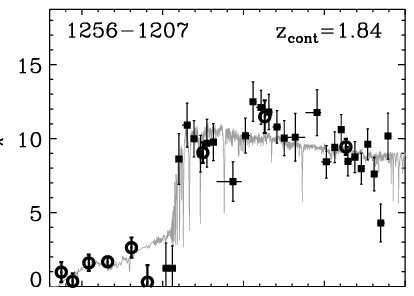}
  \includegraphics[scale=1]{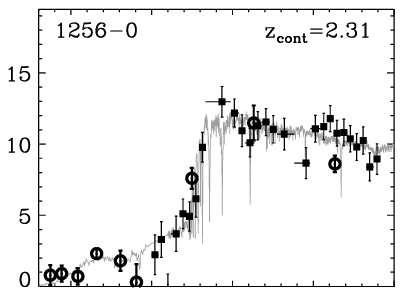}
  \includegraphics[scale=1]{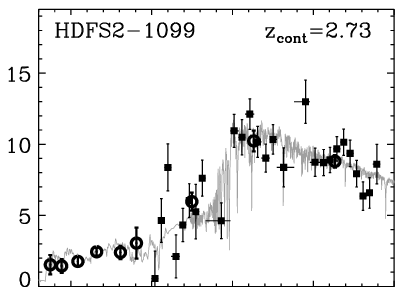}
  \includegraphics[scale=1]{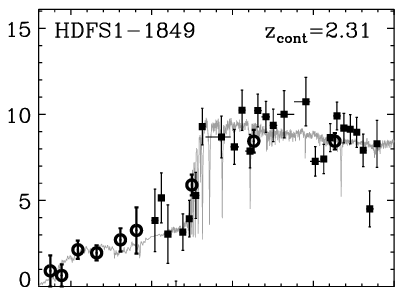}\\
  \vspace{-0.02in} \includegraphics[scale=1]{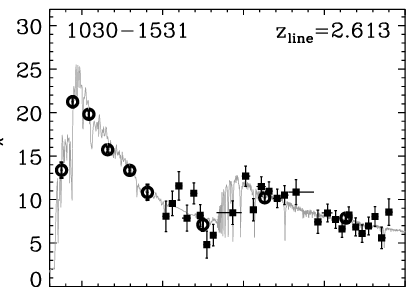}
  \includegraphics[scale=1]{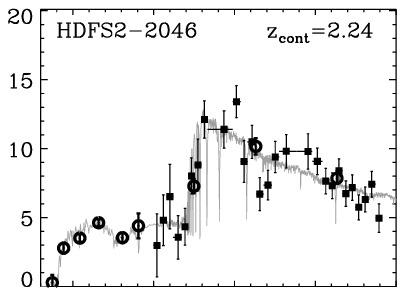}
  \includegraphics[scale=1]{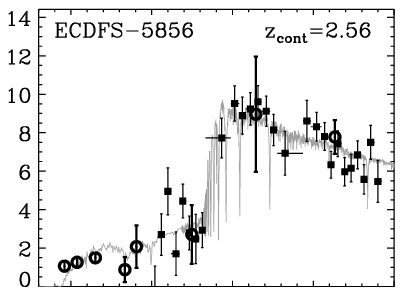}
  \includegraphics[scale=1]{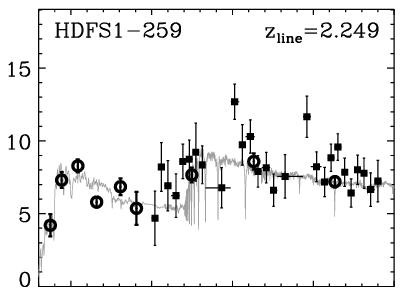}\\ \vspace{-0.02in}
  \includegraphics[scale=1]{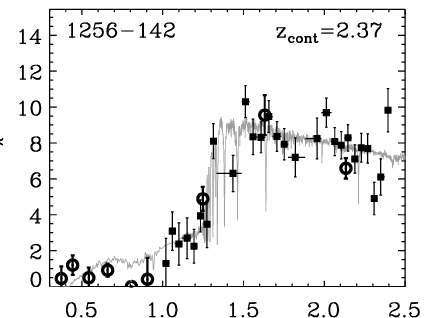}
  \includegraphics[scale=1]{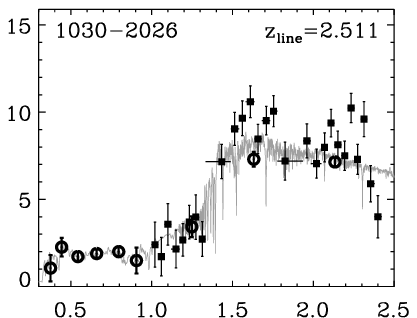}
  \includegraphics[scale=1]{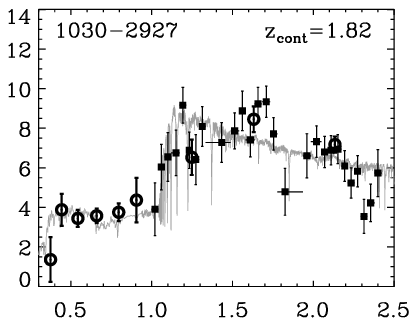}
  \includegraphics[scale=1]{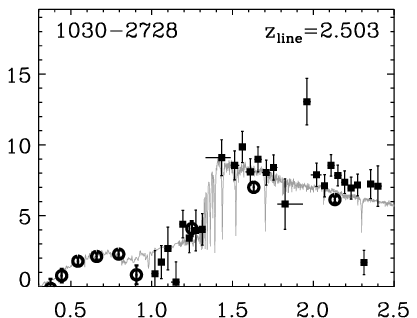}\\\vspace{0.2in}
  \figcaption{GNIRS spectra ({\it black squares}) and MUSYC broadband
  photometry ({\it open circles}) for all 36 galaxies, sorted for
  their total $K$-band magnitude starting with the brightest galaxy
  (see Table~\ref{sample}). Fluxes are given in $10^{-19}\,\rm
  ergs\,s^{-1}\,cm^{-2}\,\AA^{-1}$. Emission line fluxes are removed
  from the binned spectra, using the best-fit models to the lines. The
  photometry is not corrected for emission line contamination. The
  best-fit stellar population models to the spectra and photometry are
  shown in gray. For emission-line galaxies the redshift was fixed at
  $z_{\rm line}$ during fitting, while for the remaining galaxies $z$
  was a free parameter.\\\label{spec} } \end{figure*} }

\def\figf{
  \begin{figure*}[!t]
  \centering
      \includegraphics[scale=1]{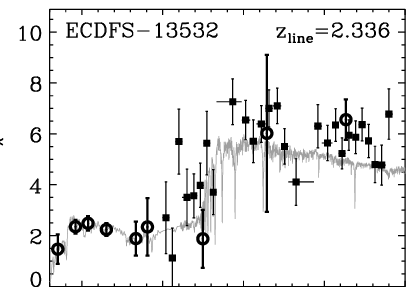}
      \includegraphics[scale=1]{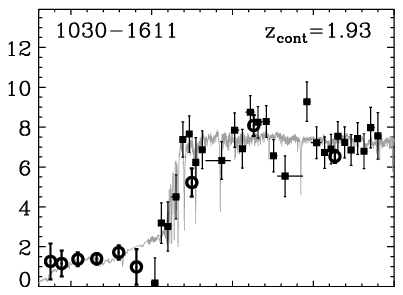}
      \includegraphics[scale=1]{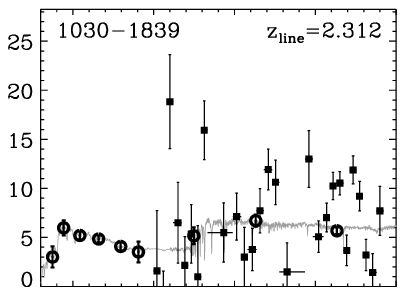}
      \includegraphics[scale=1]{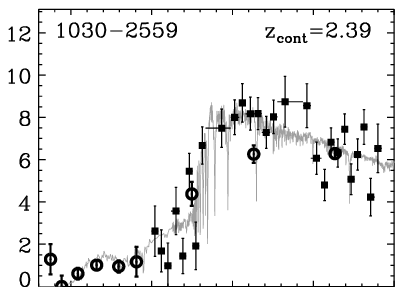}\\
      \vspace{-0.02in}
      \includegraphics[scale=1]{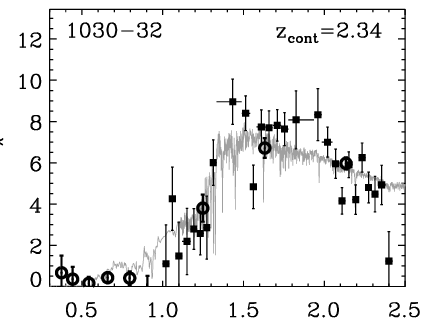}
      \includegraphics[scale=1]{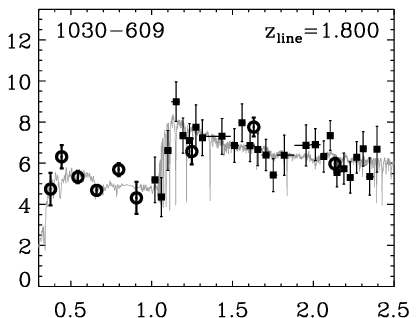}
      \includegraphics[scale=1]{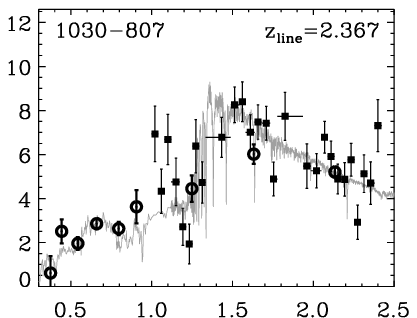}
      \includegraphics[scale=1]{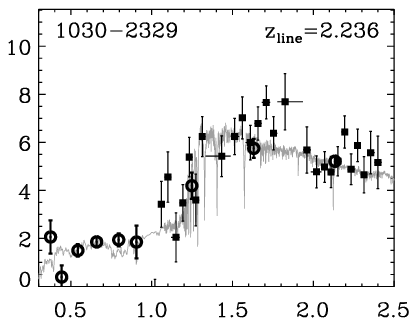}\\\vspace{0.2in} 
  \figcaption{{\it Continued}}
  \end{figure*}
}

\def\figg{
  \begin{figure}[!t]
  \centering
  \includegraphics[scale=\sonec]{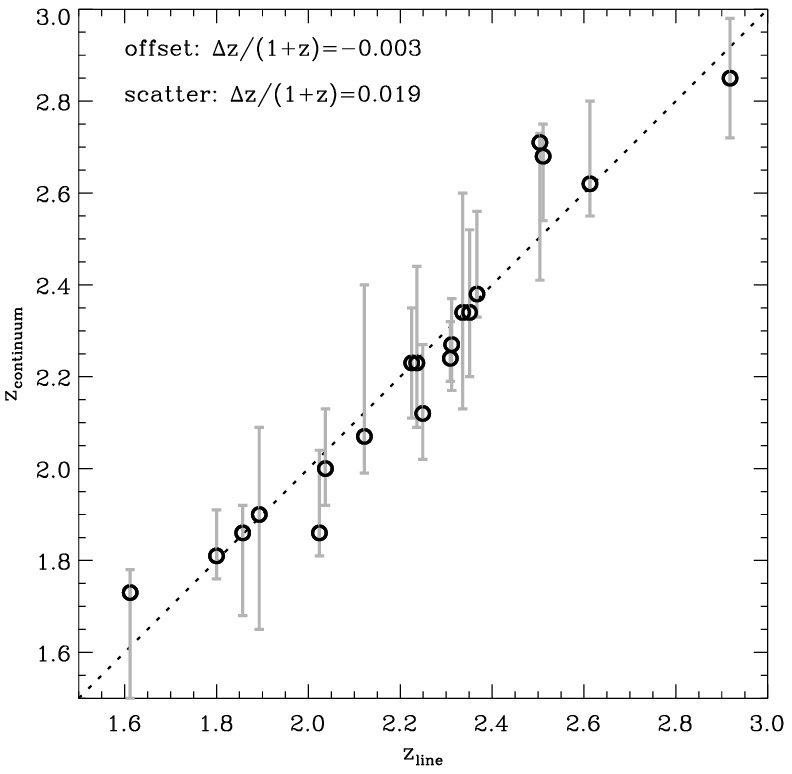}
  \figcaption{In this figure we illustrate the accuracy of continuum
    redshifts, by deriving $z_{\rm cont}$ for the 19 emission-line
    galaxies in the sample. The scatter and the systematic offset in
    \dzs\ are listed in the figure. The continuum redshifts are
    $\sim4-7$ times more accurate than photometric redshifts, and show
    no significant systematic offset. Galaxies without emission lines
    generally have larger breaks, so their $z_{\rm cont}$ may even be
    more accurate. \label{ztest}}
  \end{figure}
}

\def\figh{
  \begin{figure}[!t]
  \centering
  \includegraphics[scale=\sonec]{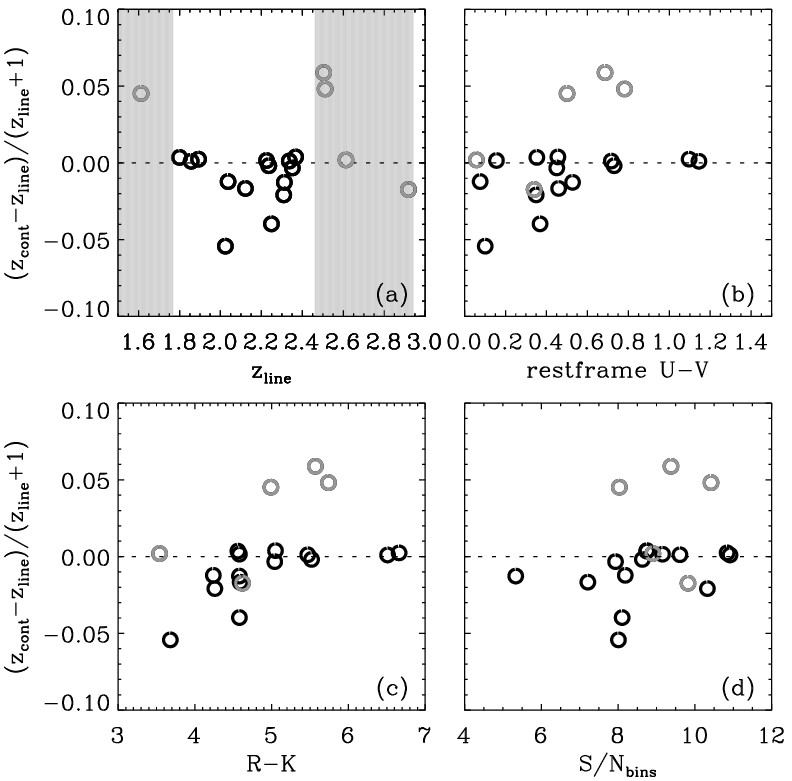}
  \figcaption{In this diagram we examine the cause of errors in
    continuum redshifts. As the modeling is mainly driven by the
    optical break, we expect and indeed find less accurate continuum
    redshifts for galaxies for which a large part of the break falls
    between atmospheric windows or outside the spectrum (panel $a$,
    {\it shaded regions}). These galaxies are indicated by gray
    symbols in all panels. In panels $b$ and $c$ we show that for
    galaxies with bluer SEDs, and thus weaker optical breaks, the
    continuum redshifts are less accurate. There is no clear
    correlation with the S/N of the spectrum in panel
    $d$. \label{ztest2}}
\end{figure} 
}

\def\figi{
  \begin{figure}[!t]
  \centering
  \includegraphics[scale=\sonec]{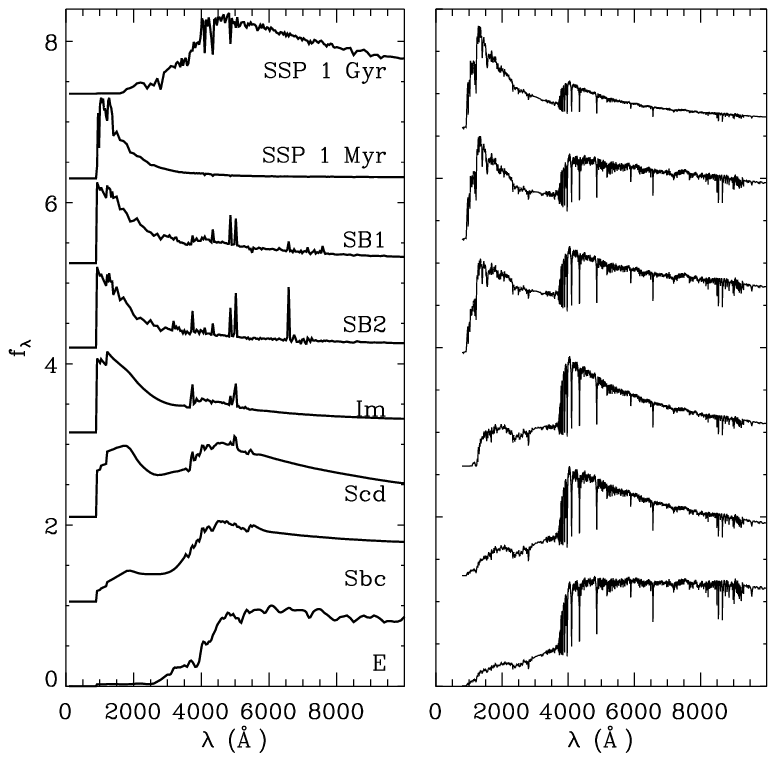}
  \figcaption{{\it left}: Galaxy templates as used by the original
    photometric redshift code. {\it right}: Galaxy templates as
    constructed from the GNIRS sample. \label{templates}}
  \end{figure} 
}

\def\figj{
  \begin{figure}[!t] 
  \centering
  \includegraphics[scale=\sonec]{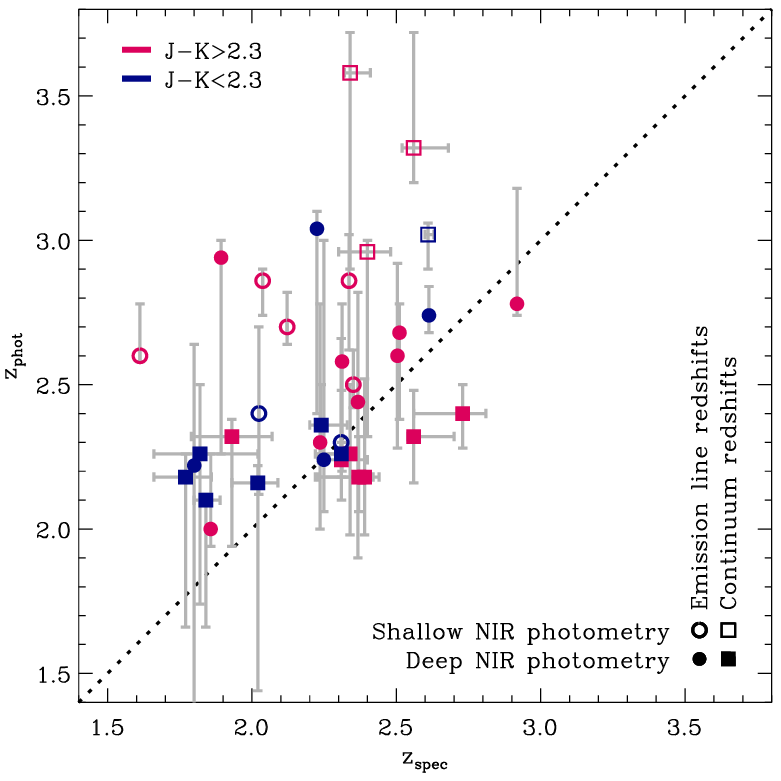}
  \figcaption{Spectroscopic vs. photometric redshifts. The scatter and
    offset in \dzs\ for the full sample are 0.13 and 0.08
    respectively. The squares and circles represent the continuum and
    emission line redshifts respectively, for galaxies in the deep
    ({\it filled symbols}) and wide ({\it open symbols}) MUSYC
    fields. DRGs \citep[$J-K>2.3$,][]{fr03} are indicated in red and
    non-DRGs are shown in blue. Corresponding systematic and random
    errors are given in Table~\ref{tab_dz}.  \label{z_comp}}
  \end{figure} 
}

\def\figk{
  \begin{figure}[!t]
  \centering
  \includegraphics[scale=\sonec]{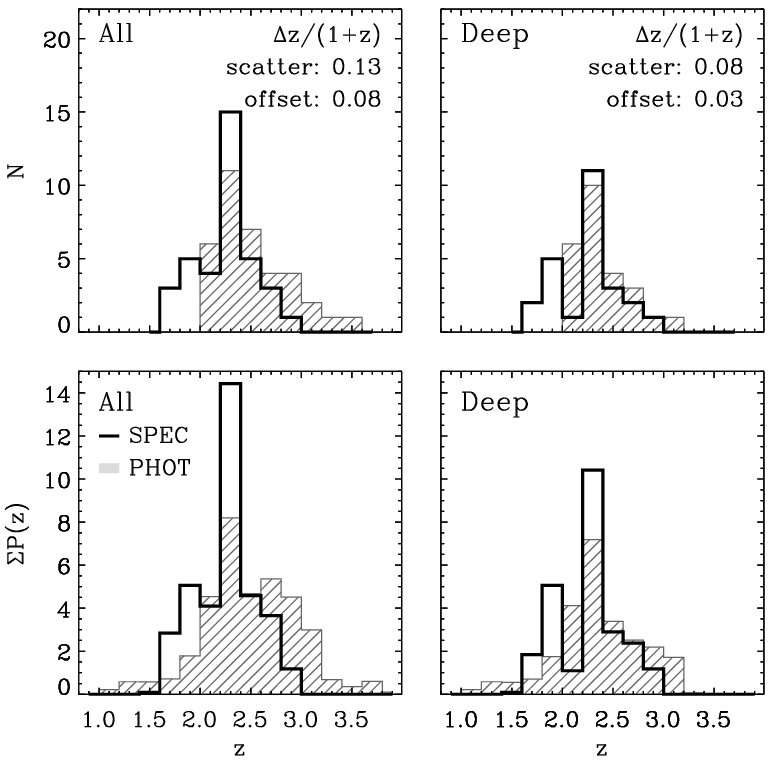}\\
  \includegraphics[scale=\sonec]{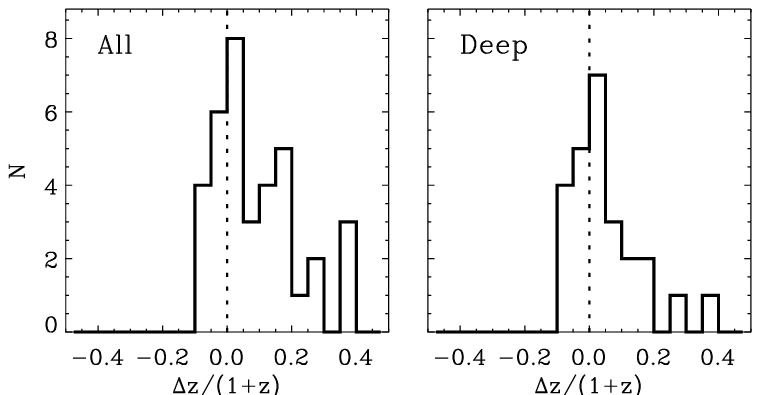}
  \figcaption{Spectroscopic ({\it black solid histograms}) and
    photometric ({\it diagonally hatched histograms}) redshift
    distributions ({\it top panels}) and their summed probability
    distributions ({\it middle panels}) for the full spectroscopic
    sample ({\it left panels}), and for those galaxies with deep NIR
    photometry ({\it right panels}). The scatter and systematic offset
    in \dzs\ are given in the top right corner in the top two
    panels. The lower panels show the distribution of \dzs\ for the
    full ({\it left}) and deep ({\it right}) sample
    respectively.\label{z_hist}}
  \end{figure} 
}

\def\figl{
  \begin{figure*}[!t] 
  \begin{flushleft}
    \includegraphics[scale=1]{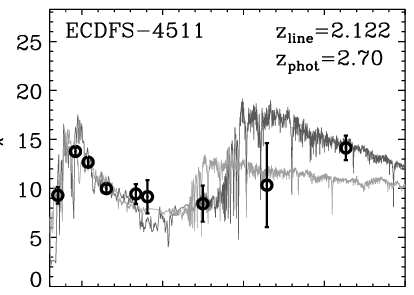}
    \includegraphics[scale=1]{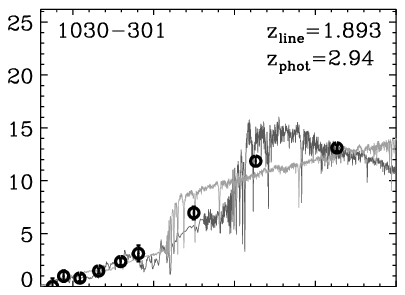}
    \includegraphics[scale=1]{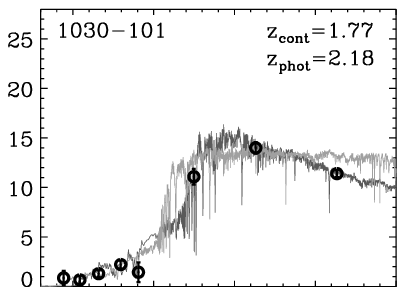}
    \includegraphics[scale=1]{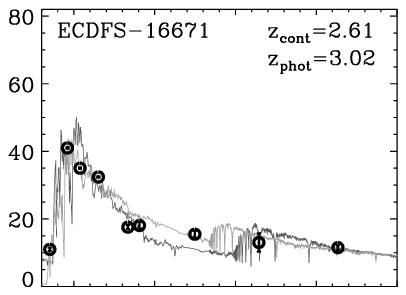}\\
    \vspace{-0.02in} \includegraphics[scale=1]{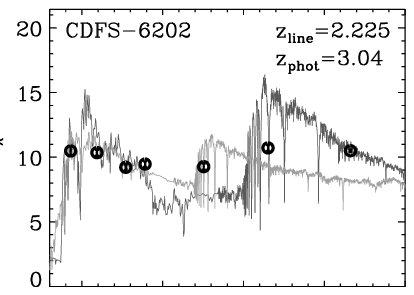}
    \includegraphics[scale=1]{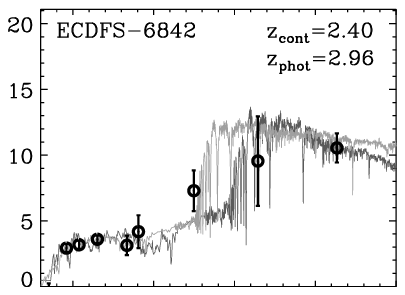}
    \includegraphics[scale=1]{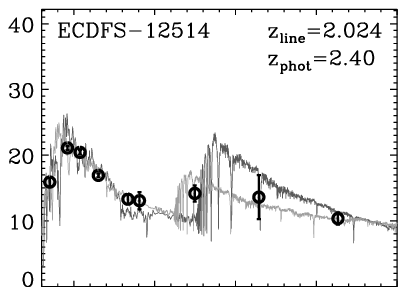}
    \includegraphics[scale=1]{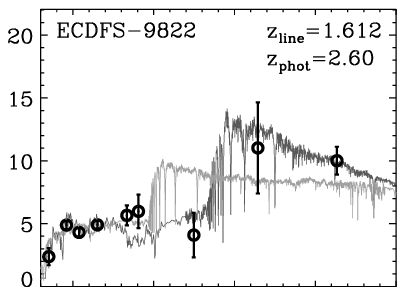}\\
    \vspace{-0.02in} \includegraphics[scale=1]{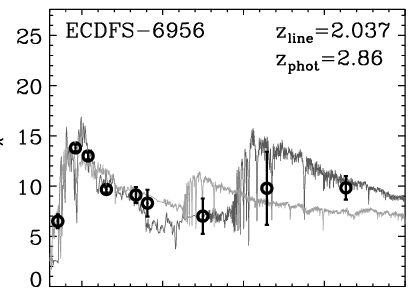}
    \includegraphics[scale=1]{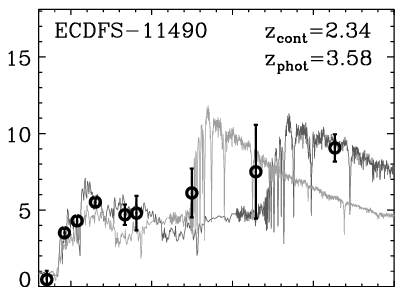}
    \includegraphics[scale=1]{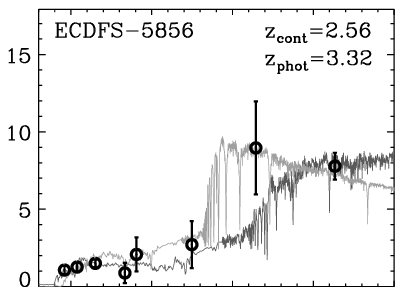}
    \includegraphics[scale=1]{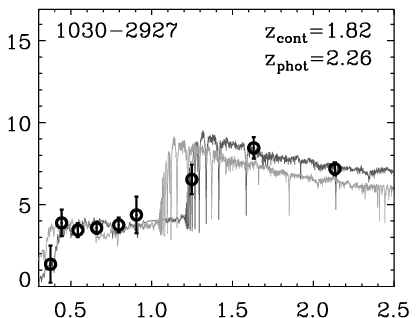}\\
    \vspace{-0.02in} \includegraphics[scale=1]{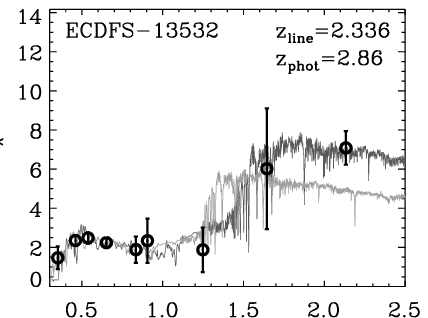}
    \includegraphics[scale=1]{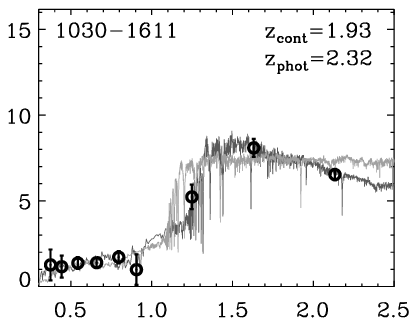}
    \includegraphics[scale=1]{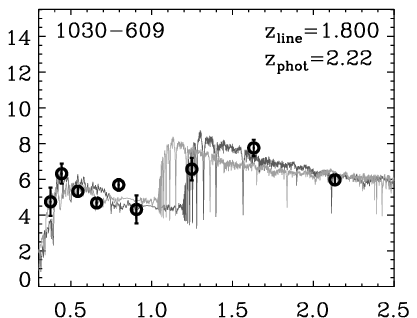}
    \end{flushleft}
  \figcaption{Broadband photometry of all galaxies for which the
    photometric and spectroscopic redshifts are substantially
    different (\dzs$>0.1$). The dark gray fit shows the best-fit
    stellar population model to just the broadband photometry. The
    best-fit stellar population model to the spectrum in combination
    with the optical photometry is presented by the light gray
    fit. Both the photometric and spectroscopic (emission-line or
    continuum) redshifts are given in the panels.\label{phot}}
\end{figure*} 
}

\def\figm{
  \begin{figure*}[!t]
  \centering
  \includegraphics[scale=\stwoc]{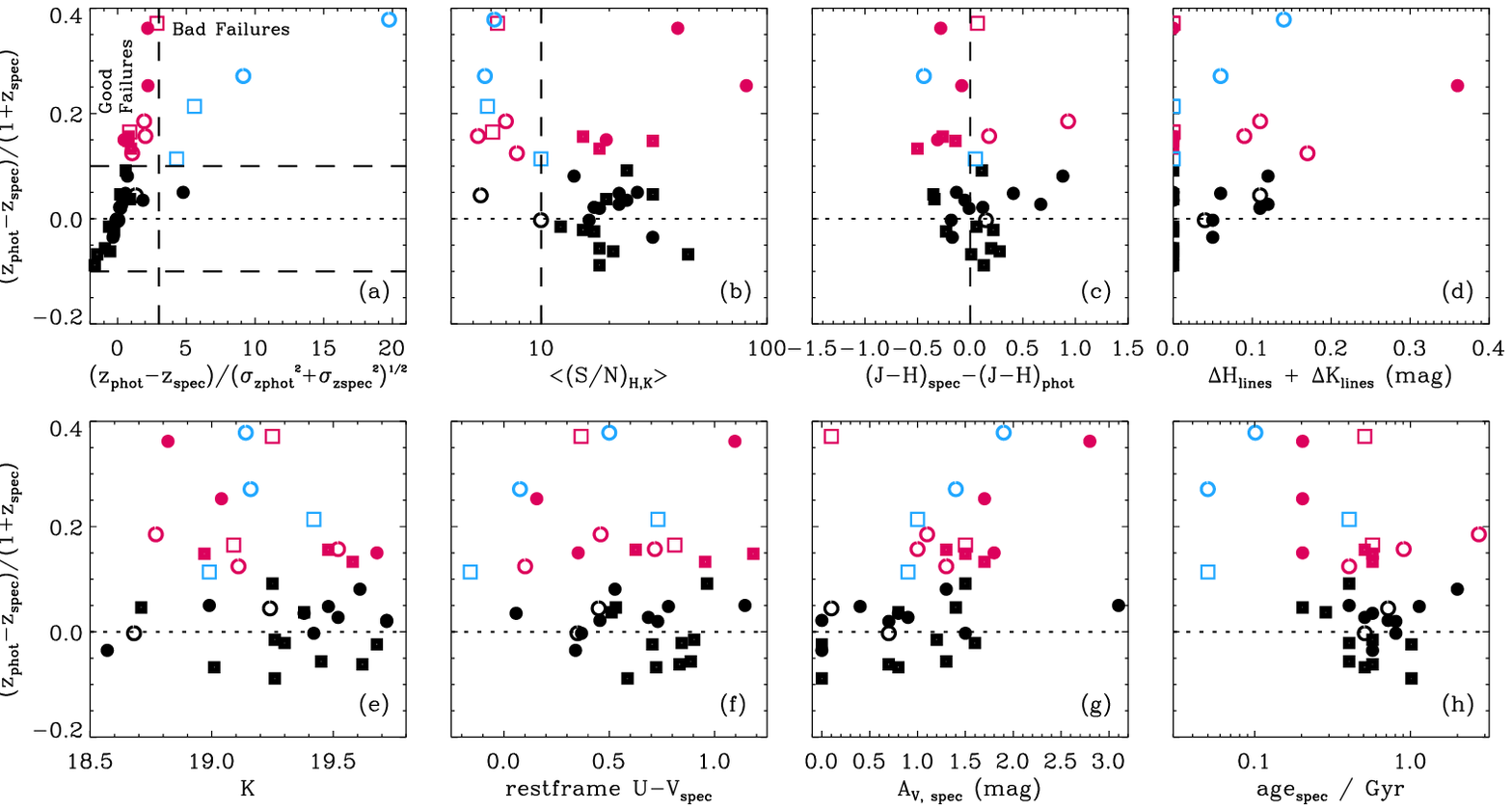}
  \figcaption{In this figure we examine systematics and catastrophic
    failures in \dzs. Squares and circles represent galaxies with
    continuum and emission-line redshifts respectively, in the wide
    ({\it open symbols}) and deep ({\it filled symbols}) MUSYC
    surveys. Galaxies with \dzs$<0.1$ are represented by the black
    symbols. In panel $a$ we classify failures (\dzs$>0.1$) such that
    galaxies for which \zs\ is within 3$\sigma$ consistent with \zp\
    are defined as ``good failures'' ({\it red}), while galaxies that
    do not meet this criterion are defined as ``bad failures'' ({\it
      blue}).  Furthermore, we relate \dz\ with the S/N of the NIR
    photometry ($b$), the difference between the photometric and
    spectroscopic $J-H$ colors ($c$), the contamination of the
    broadband photometry by emission lines ($d$), total $K$-band
    magnitude ($e$), rest-frame $U-V$ color ($f$), and the stellar
    population properties $A_V$ ($g$) and age ($h$).  This figure
    illustrates that \dz\ is generally larger for galaxies with
    shallow NIR photometry, and mainly correlates with SED
    type.  \label{sys}}
  \end{figure*} 
}

\def\fign{
  \begin{figure}[!t] 
    \begin{center}
      \includegraphics[scale=1.0]{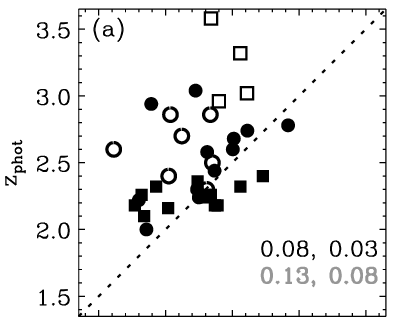}\hspace{-0.2in}
      \includegraphics[scale=1.0]{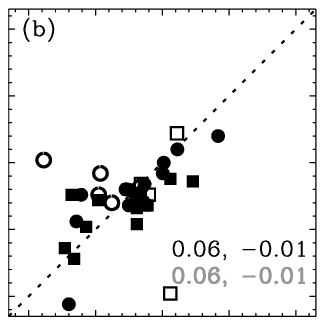}\\\vspace{-0.2in}
      \includegraphics[scale=1.0]{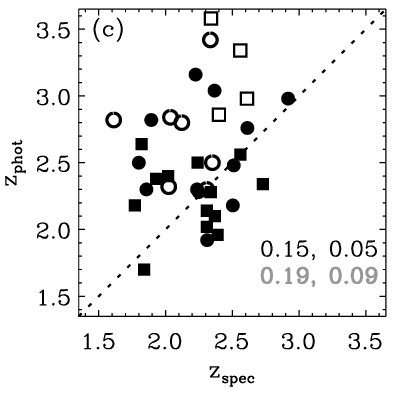}\hspace{-0.2in}
      \includegraphics[scale=1.0]{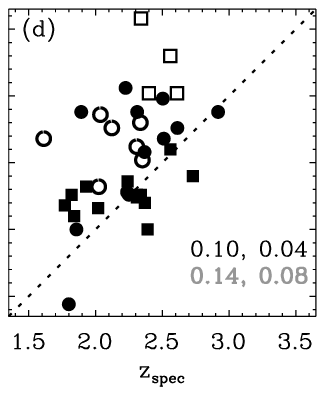}
    \end{center}
    \figcaption{In this figure we test the different causes for
      scatter and systematics in \dzs. In panels $a$ we show the
      original \zp\ and in panels $b$ we use the new templates as
      shown in Figure~\ref{templates} to derive photometric
      redshifts. In panel $c$ we have repeated the original fitting
      procedure, replacing the NIR photometry by the spectroscopic
      $J$, $H$ and $K$ fluxes. In panel $d$ we remove the emission
      line fluxes from the broadband photometry before applying the
      photometric redshift procedure. Circles and squares represent
      galaxies with emission line and continuum redshifts
      respectively. For galaxies with deep NIR photometry the symbols
      are filled. The scatter and systematic in \dzs\ are given in
      black for the high S/N galaxies and in gray for the total
      sample. This figure illustrates that the systematic offset in
      \zp\ can almost completely be removed and the absolute errors
      reduced by using well calibrated templates.\label{newz}}
  \end{figure} 
}

\def\figo{
  \begin{figure*}[!t] 
    \centering 
    \includegraphics[scale=1.05]{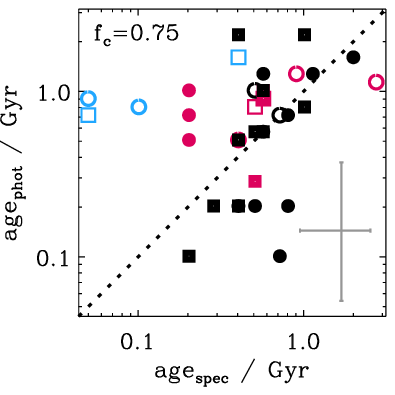}
    \includegraphics[scale=1.05]{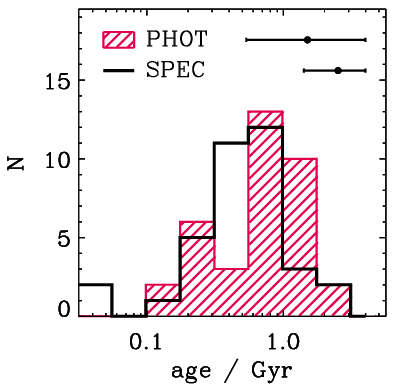}
    \includegraphics[scale=1.05]{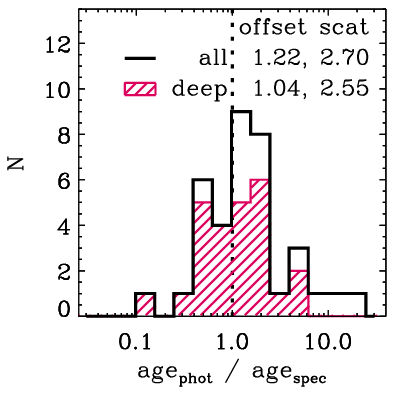}
    \includegraphics[scale=1.05]{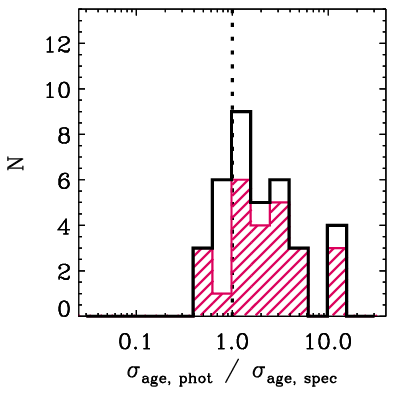}\\\
    \includegraphics[scale=1.05]{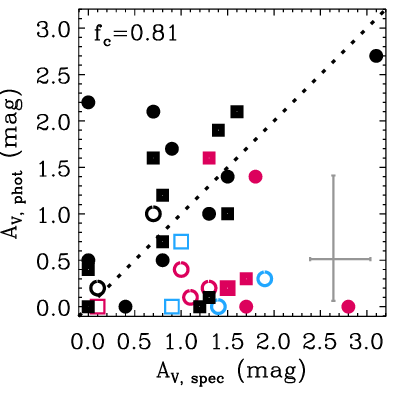}
    \includegraphics[scale=1.05]{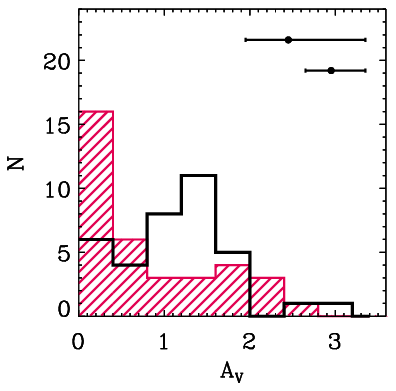}
    \includegraphics[scale=1.05]{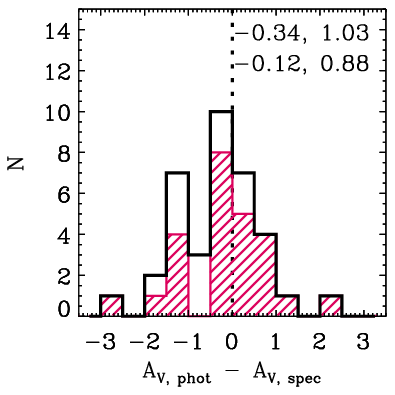}
    \includegraphics[scale=1.05]{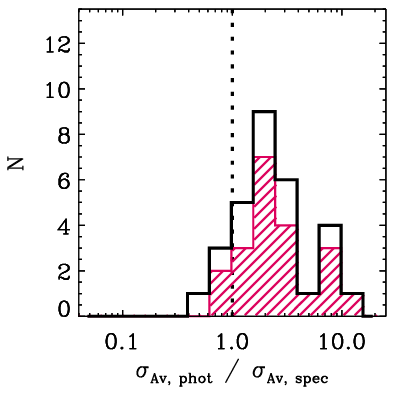}\\
    \includegraphics[scale=1.05]{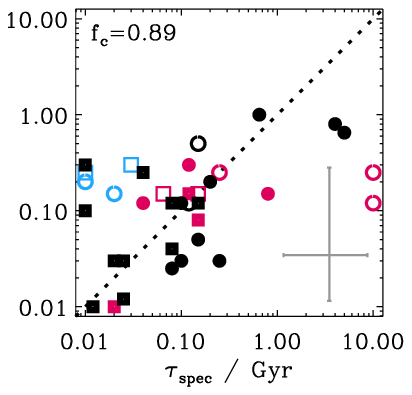}
    \includegraphics[scale=1.05]{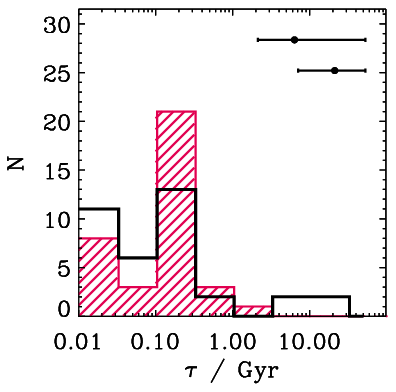}
    \includegraphics[scale=1.05]{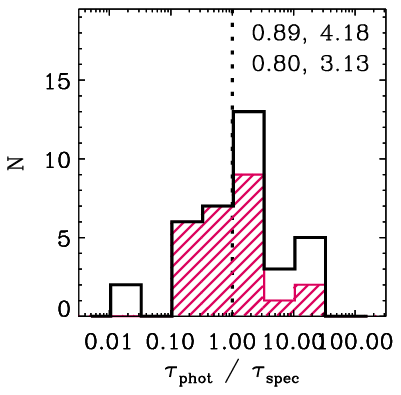}
    \includegraphics[scale=1.05]{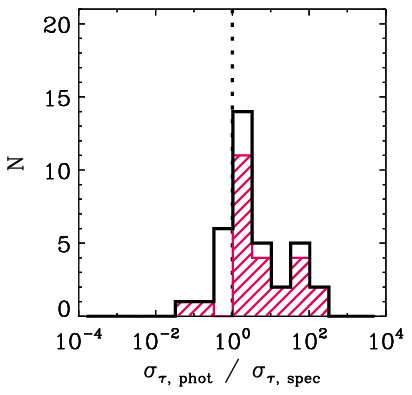}\\
    \includegraphics[scale=1.05]{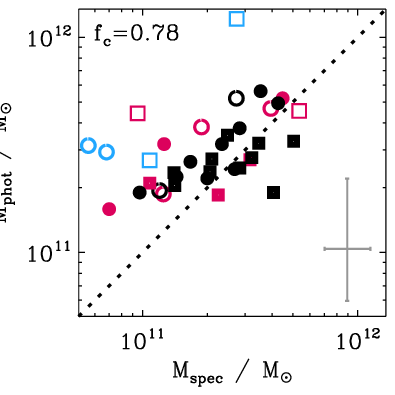}
    \includegraphics[scale=1.05]{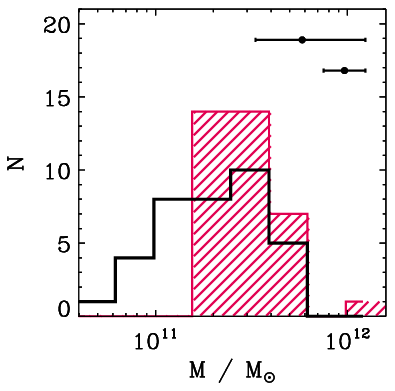}
    \includegraphics[scale=1.05]{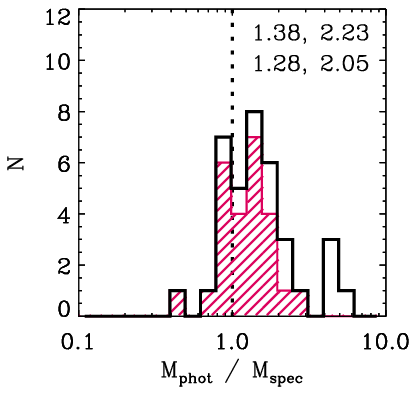}
    \includegraphics[scale=1.05]{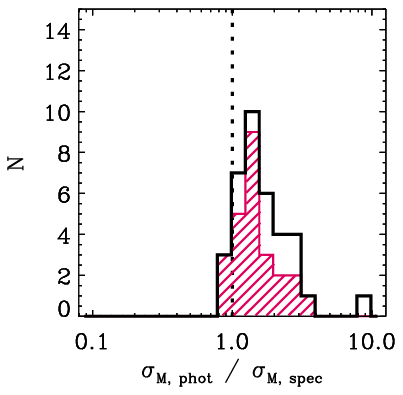}\\
    \includegraphics[scale=1.05]{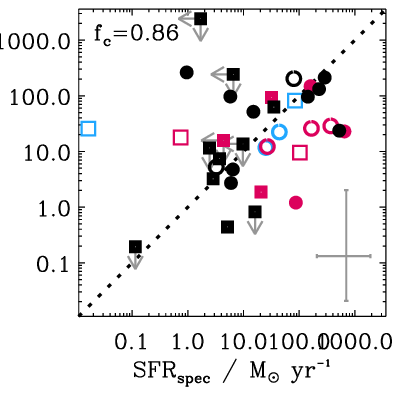}
    \includegraphics[scale=1.05]{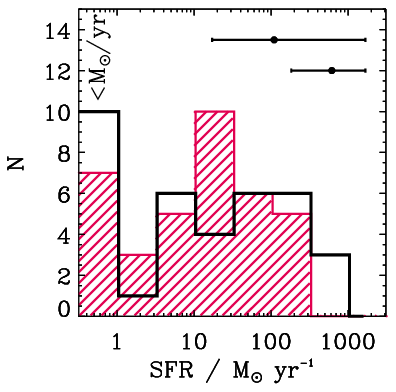}
    \includegraphics[scale=1.05]{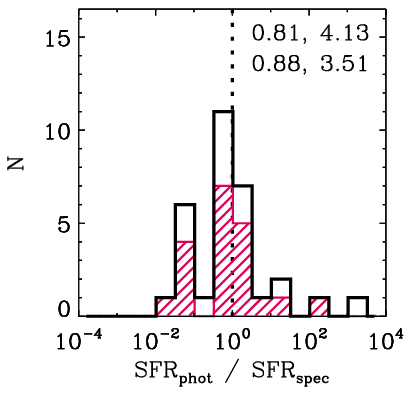}
    \includegraphics[scale=1.05]{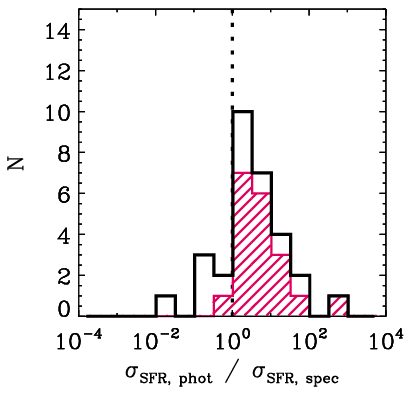}
  \figcaption{Comparison of the photometric and spectroscopic
      properties age (since the onset of star formation), $A_V$,
      $\tau$, stellar mass, SFR, $M/L_V$ ratio, rest-frame V
      magnitude, and rest-frame U-V color. The left panels show the
      direct comparison for the individual galaxies with the symbols
      similar as in Figure~\ref{sys}. The fraction ($F_{\rm c}$) of
      galaxies for which the photometric and spectroscopic properties
      are consistent within 1$\sigma$ is given in the top left corner,
      and the typical error in the bottom right. The dotted diagonal
      lines indicate where the spectroscopic properties are equal to
      the photometric properties. The panels in the second column show
      both distributions. The typical errors for the photometric and
      spectroscopic properties are given in the top right corner. The
      distribution of the difference between the photometric and
      spectroscopic properties is presented in the panels in the third
      column. The offset (biweight mean) and scatter (normalized
      biweight mean absolute deviation) of the distribution is given
      for each property for the total and deep sample. In the right
      panels we present the distribution of the ratio of the
      photometric and spectroscopic confidence intervals for the
      individual galaxies. The black histograms represent all
      galaxies, and the colored, diagonally hatched areas correspond
      to the galaxies with deep NIR photometry. This figure
      illustrates that the stellar population properties such as the
      age, $A_V$, SFR, and $\tau$ are generally very poorly
      constrained with broadband data alone. Interestingly, stellar
      masses and $M/L_V$ are among the most stable
      properties. \label{prop}}
  \end{figure*} 
}

\def\figp{
  \begin{figure*}[!t]
  \centering
      \includegraphics[scale=1.05]{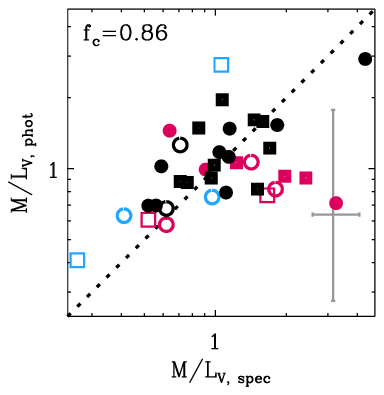}
      \includegraphics[scale=1.05]{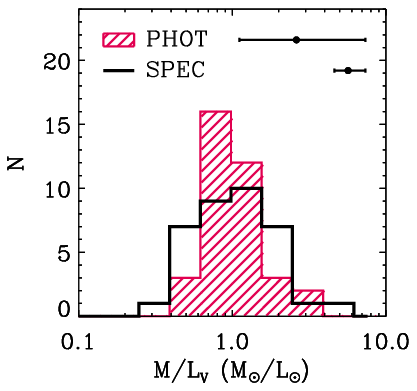}
      \includegraphics[scale=1.05]{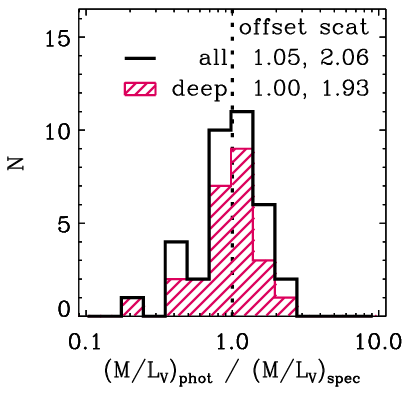}
      \includegraphics[scale=1.05]{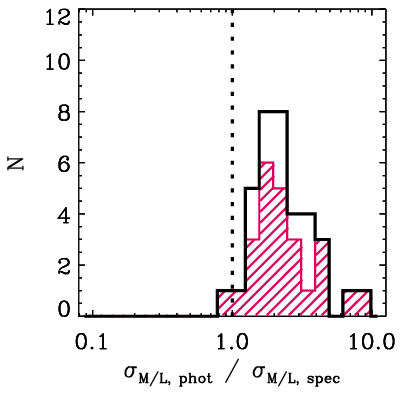}\\
      \includegraphics[scale=1.05]{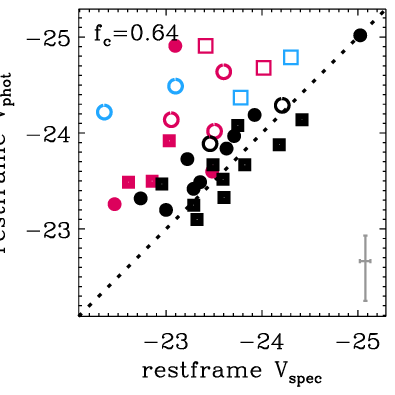}
      \includegraphics[scale=1.05]{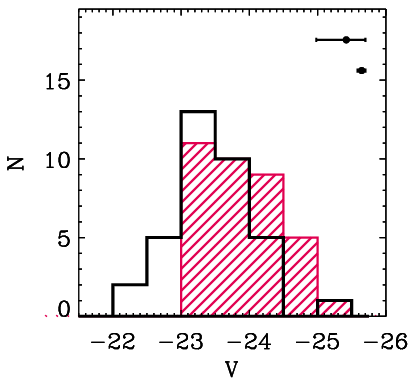}
      \includegraphics[scale=1.05]{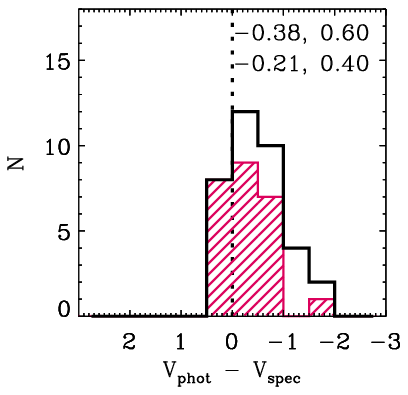}
      \includegraphics[scale=1.05]{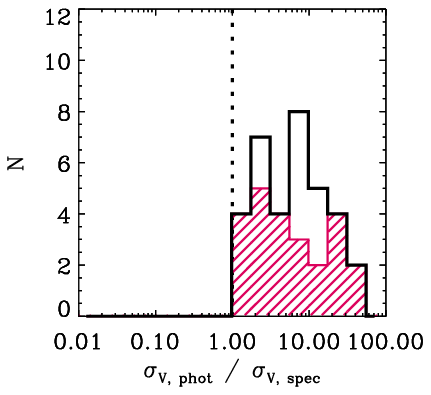}\\
      \includegraphics[scale=1.05]{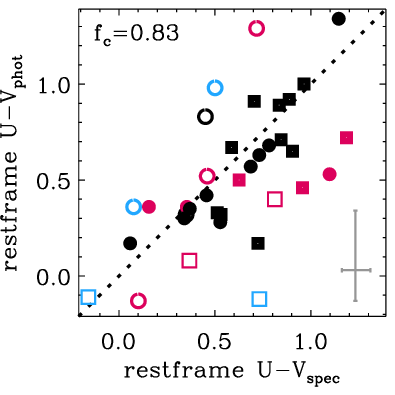}
      \includegraphics[scale=1.05]{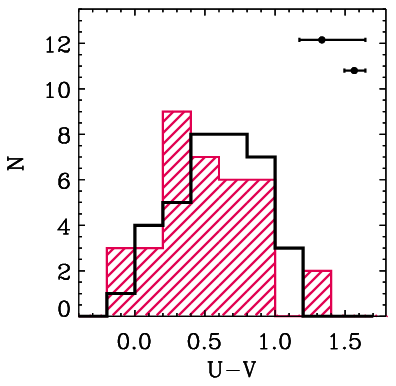}
      \includegraphics[scale=1.05]{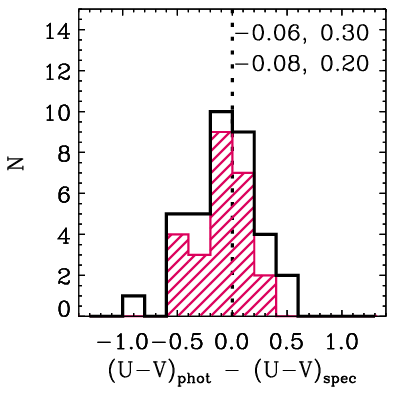}
      \includegraphics[scale=1.05]{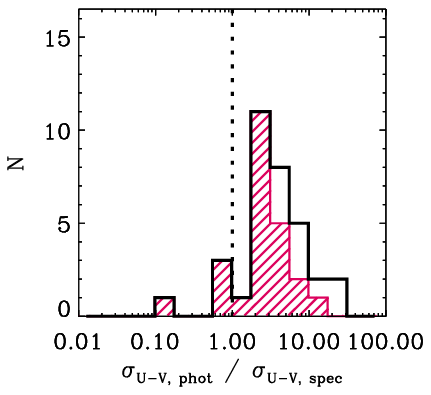}
  \figcaption{{\it Continued}}
  \end{figure*} 
}

\def\figq{
  \begin{figure*}[!t] 
  \centering
    \includegraphics[scale=1.05]{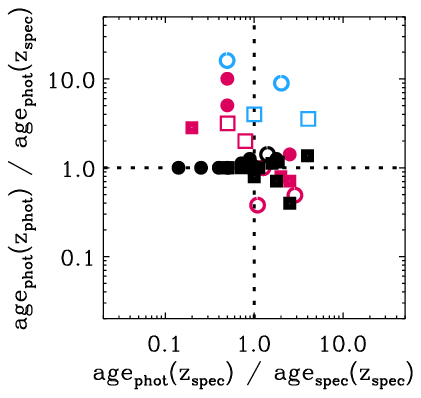}%\hspace{0.05in}
    \includegraphics[scale=1.05]{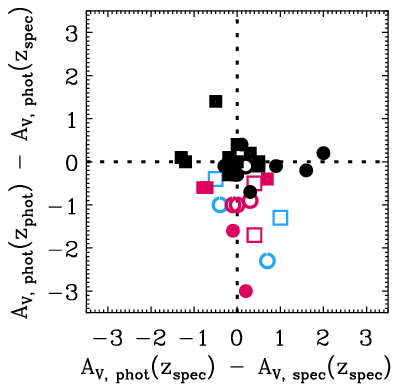}%\hspace{0.05in}
    \includegraphics[scale=1.05]{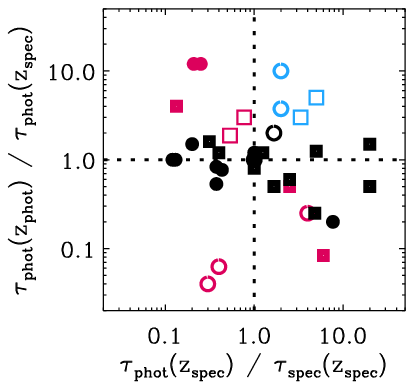}%\hspace{0.05in}
    \includegraphics[scale=1.05]{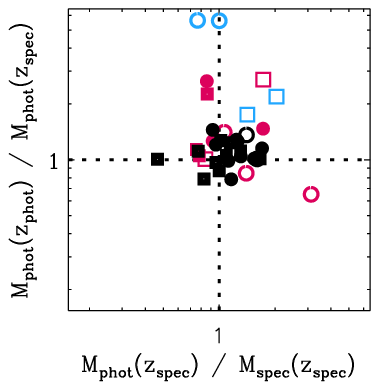}\\
    \includegraphics[scale=1.05]{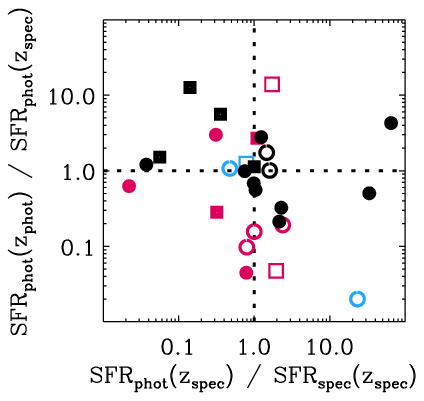}%\hspace{0.05in}
    \includegraphics[scale=1.05]{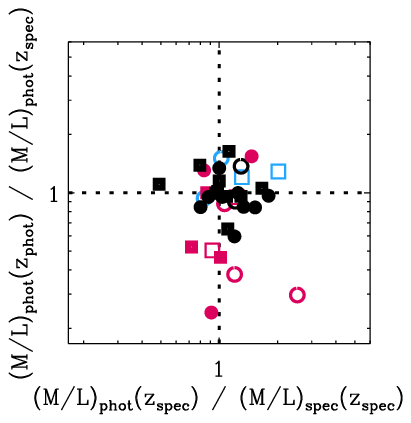}%\hspace{0.05in}
    \includegraphics[scale=1.05]{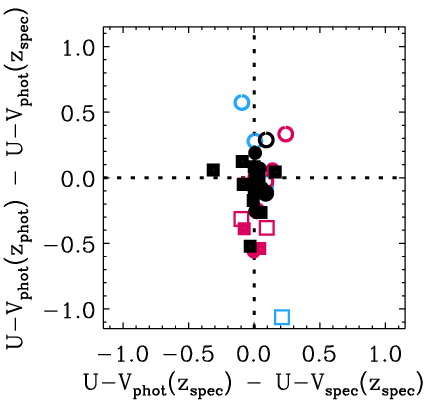}%\hspace{0.05in}
    \includegraphics[scale=1.05]{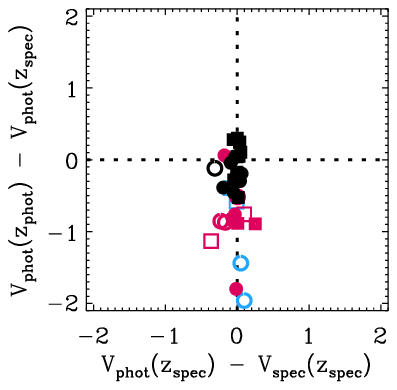}
  \figcaption{In this figure we examine the influence of both the
    better constrained redshift and the higher resolution spectral
    shape on the stellar population properties (see also
    Table~\ref{tab_scatter}). The y-axes present the difference
    between best-fit properties by including spectroscopic
    redshifts. We derived this improvement by fitting the photometry
    replacing \zp\ by \zs. On the x-axes we show the additional
    improvement due to the higher resolution spectral shape. This
    improvement is the difference between the best-fit values for
    fitting the photometry with the redshift fixed to $z_{\rm spec}$,
    in- and excluding the NIR spectrum. Symbols are similar as in
    Figure~\ref{sys}. \label{what}}
  \end{figure*} 
}

\def\figr{
  \begin{figure*}[!t] 
  \centering
    \includegraphics[scale=1.05]{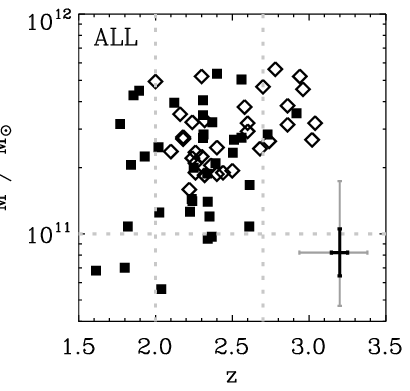}\hspace{0.05in}
    \includegraphics[scale=1.05]{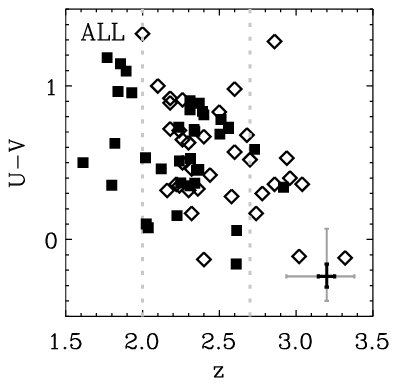}\hspace{0.05in}
    \includegraphics[scale=1.05]{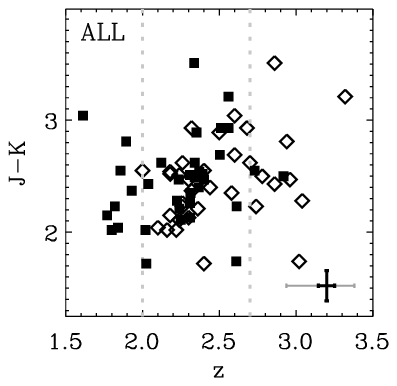}\hspace{0.05in}
    \includegraphics[scale=1.05]{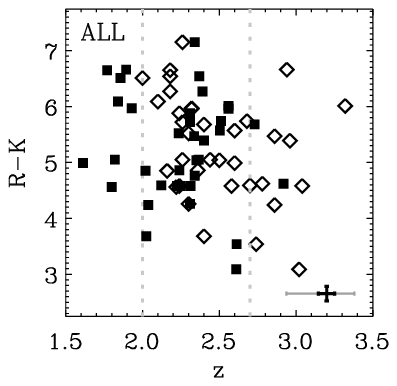}\\
    \includegraphics[scale=1.05]{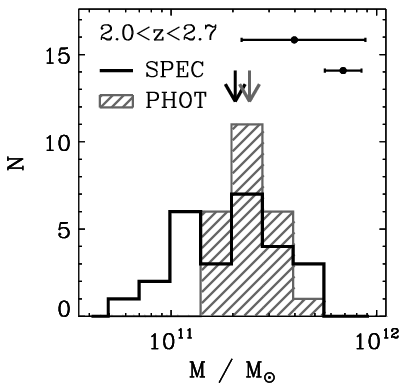}\hspace{0.05in}
    \includegraphics[scale=1.05]{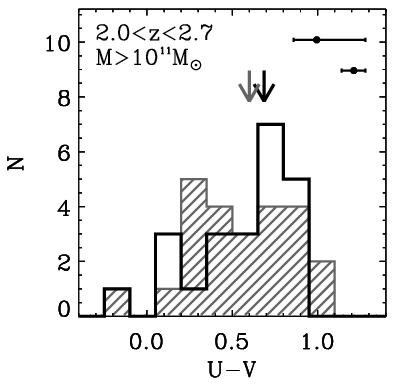}\hspace{0.05in}
    \includegraphics[scale=1.05]{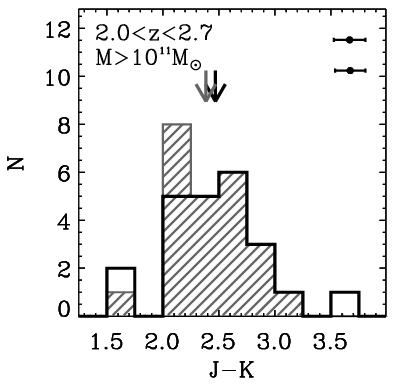}\hspace{0.05in}
    \includegraphics[scale=1.05]{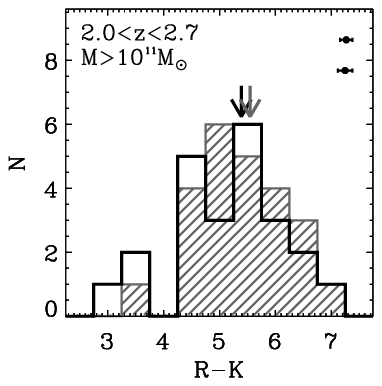}
  \figcaption{Comparison of photometrically ({\it open diamonds}) and
    spectroscopically ({\it filled squares}) derived properties of the
    full spectroscopic sample. The top panels show stellar mass,
    rest-frame $U-V$ color, observed $J-K$ and $R-K$ color versus
    redshift. Hence all galaxies are plotted twice. The $J-K$ and
    $R-K$ colors are in both cases photometric, while for the other
    properties ($z$, $U-V$ and stellar mass) we adopt the
    spectroscopic or photometric values. The bottom panels show for
    both the distribution of the stellar mass, rest-frame $U-V$ color,
    observed $J-K$ and $R-K$ color for galaxies in the same redshift
    range, and for the latter three, the same mass range. The arrows
    indicate the median values of the distributions. The typical
    errorbars are shown in the bottom and top right of the top and
    bottom panels respectively. The barely changed distributions of
    $J-K$ and $R-K$ imply that the spectroscopy supports the previous
    finding that red galaxies dominate the high mass end at
    $2<z<3$.\label{impl}}
  \end{figure*} 
}

\def\figs{
  \begin{figure}[!t]
  \centering
  \includegraphics[scale=\sonec]{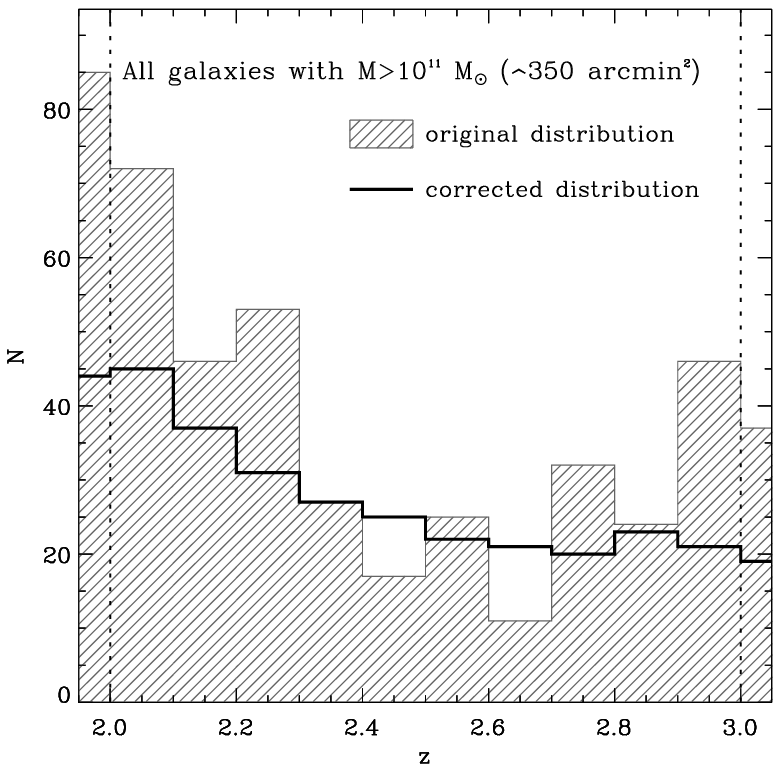} 
  \figcaption{The original and corrected number of galaxies with
    $M>10^{11} M_{\odot}$ vs. redshift in the deep MUSYC fields. The
    original distribution is based on \zp\ and photometric masses. The
    corrected distribution is estimated using the coupled
    distributions of \dz\ and $M_{\rm spec}/M_{\rm phot}$, as
    explained in the text. This figure illustrates that the number of
    massive galaxies with $M>10^{11} M_{\odot}$ in the range $2<z<3$
    is likely lower than previously derived by \cite{vd06} using only
    photometric information. \label{model_n}}
  \end{figure}
}

\def\taba{
  \begin{deluxetable*}{lrrllllrr}%[!t]
    \tabletypesize{\scriptsize} 
    \tablecaption{Sample and Observations\label{sample}}
    \tablewidth{0pt}
    \tablehead{ \colhead{} & \colhead{} & \colhead{} & \colhead{} & 
    \colhead{} & \colhead{} & \colhead{} & \colhead{Exptime} & \colhead{} \\
      \colhead{ID} & \colhead{RA} & \colhead{DEC} & \colhead{$K_s$} & 
      \colhead{$R$} & \colhead{$J-K$} & \colhead{observation dates} & 
      \colhead{min} & \colhead{ID$_{\rm KR06B+07}$\tablenotemark{a}}}
    \startdata 
    1030-32     & 10 30 41.5 &   5 19 55 & 19.68 & 26.83 & 2.62 & 2006/12/19 + 2007/03/12        & 230 &     - \\
    1030-101    & 10 30 10.1 &   5 20 11 & 18.97 & 25.63 & 2.15 & 2006/02/23                     & 120 &     - \\
    1030-301    & 10 30 50.8 &   5 20 49 & 18.82 & 25.49 & 2.81 & 2006/01/20                     &  90 &     - \\
    1030-609    & 10 30 49.6 &   5 21 50 & 19.68 & 24.23 & 2.02 & 2006/02/24                     & 130 &     - \\
    1030-807    & 10 30 20.0 &   5 22 33 & 19.72 & 24.77 & 2.40 & 2006/02/23                     & 120 &     - \\
    1030-1531   & 10 30 38.9 &   5 24 52 & 19.38 & 22.92 & 2.23 & 2006/02/25                     &  80 &     - \\
    1030-1611   & 10 30 48.4 &   5 25 03 & 19.58 & 25.55 & 2.37 & 2006/02/24                     & 120 &     - \\
    1030-1813   & 10 30 51.2 &   5 25 36 & 19.01 & 24.97 & 2.93 & 2006/01/20                     &  80 &     - \\
    1030-1839   & 10 30 45.4 &   5 30 07 & 19.61 & 24.20 & 2.35 & 2006/12/16                     &  80 &     - \\
    1030-2026   & 10 30 22.7 &   5 28 26 & 19.48 & 25.22 & 2.93 & 2006/02/22                     & 120 &     - \\
    1030-2329   & 10 30 16.2 &   5 27 32 & 19.72 & 25.24 & 2.47 & 2006/02/25                     & 120 &     - \\
    1030-2559   & 10 30 40.1 &   5 26 34 & 19.62 & 25.89 & 2.52 & 2006/02/22                     & 110 &     - \\
    1030-2728\tablenotemark{b} & 10 30 18.4 & 5 26 05 & 19.52 & 25.09 & 2.69 & 2006/01/21                     & 120 &     - \\
    1030-2927   & 10 30 43.3 &   5 29 34 & 19.48 & 24.52 & 2.23 & 2006/12/18 + 2007/03/13        & 230 &     - \\
    1256-0      & 12 54 59.6 &   1 11 30 & 19.26 & 24.98 & 2.26 & 2005/05/19+27+30  + 2006/02/24 & 305 &   151 \\
    1256-142    & 12 55 02.7 &   1 07 32 & 19.45 & 25.99 & 2.54 & 2006/02/23                     & 120 &   465 \\
    1256-519    & 12 55 08.4 &   1 06 14 & 18.99 & 25.51 & 2.55 & 2006/02/25                     &  80 &     - \\
    1256-1207   & 12 55 19.7 &   1 12 46 & 19.25 & 25.34 & 2.04 & 2006/02/25                     &  80 &     - \\
    1256-1967   & 12 55 25.8 &   1 03 25 & 18.71 & 23.55 & 2.02 & 2005/05/18 + 2006/01/18        & 240 &  2889 \\
    HDFS1-259   & 22 33 11.2 & -60 40 47 & 19.42 & 24.00 & 2.11 & 2006/12/17-18                  & 140 &     - \\
    HDFS1-1849  & 22 33 37.9 & -60 33 15 & 19.30 & 25.18 & 2.51 & 2004/09/06                     & 115 &     - \\
    HDFS2-509   & 22 31 23.1 & -60 39 08 & 18.57 & 23.20 & 2.50 & 2005/05/16+19                  & 235 &     - \\
    HDFS2-1099  & 22 32 03.2 & -60 36 13 & 19.26 & 24.94 & 2.55 & 2006/12/19                     & 120 &     - \\
    HDFS2-2046  & 22 32 30.8 & -60 32 44 & 19.38 & 24.24 & 2.21 & 2005/05/20                     & 125 &     - \\
    ECDFS-4454  &  3 32 11.5 & -27 55 23 & 19.24 & 24.28 & 2.89 & 2006/01/18                     & 100 &  3662 \\ 
    ECDFS-4511  &  3 32 43.2 & -27 55 15 & 18.77 & 23.36 & 2.62 & 2006/01/21 + 20006/02/25       & 190 &  3694 \\ 
    ECDFS-4713  &  3 31 52.5 & -27 54 48 & 18.68 & 22.94 & 2.13 & 2006/02/22                     &  60 &  3896 \\ 
    ECDFS-5856  &  3 32 13.3 & -27 52 26 & 19.42 & 25.42 & 3.21 & 2006/01/19                     & 120 &  4937 \\ 
    ECDFS-6842  &  3 31 51.3 & -27 50 56 & 19.09 & 24.47 & 2.47 & 2006/12/19 + 2007/03/11+14     & 210 &     - \\ 
    ECDFS-6956  &  3 32 02.5 & -27 50 46 & 19.16 & 23.40 & 2.43 & 2006/01/20 + 2006/02/24        & 150 &  5754 \\ 
    ECDFS-9822  &  3 31 33.9 & -27 46 03 & 19.14 & 24.13 & 3.04 & 2006/12/17                     & 120 &    -  \\ 
    ECDFS-11490 &  3 32 45.0 & -27 43 09 & 19.25 & 24.01 & 2.49 & 2006/01/20+21                  & 190 &  9510 \\ 
    ECDFS-12514 &  3 31 39.5 & -27 41 20 & 19.11 & 22.79 & 1.72 & 2006/02/23                     &  90 & 10525 \\ 
    ECDFS-13532 &  3 31 54.8 & -27 39 23 & 19.52 & 24.98 & 3.51 & 2006/12/18                     & 160 &     - \\ 
    ECDFS-16671 &  3 31 58.9 & -27 35 16 & 18.99 & 22.08 & 1.74 & 2006/12/16                     &  60 &     - \\ 
    CDFS-6202   &  3 32 31.5 & -27 46 23 & 19.04 & 23.62 & 2.28 & 2004/09/02+03                  &  90 &  6036 \\ 
    \enddata
    \tablenotetext{a}{ID numbers in \cite{kr06b,kr07}}
    \tablenotetext{b}{The spectroscopic redshift of this galaxy has first been confirmed using $K$-band spectroscopy with NIRSPEC on Keck, in 2005 January.} 
  \end{deluxetable*}
}

\def\tabb{
  \renewcommand\baselinestretch{1.4}
  \begin{deluxetable*}{lllrlllllr}%[!t]
    \tabletypesize{\scriptsize} 
    \tablecaption{Spectroscopic modeling results\label{tab_spec}}
    \tablewidth{0pt}
    \tablehead{ \colhead{} & \colhead{} & \colhead{age} & 
      \colhead{$\tau$} & \colhead{$A_V$} & \colhead{$M$} & \colhead{SFR} & 
      \colhead{$M/L_V$} & \colhead{} & \colhead{} \\
      \colhead{ID} & \colhead{$z$\tablenotemark{a}} & \colhead{Gyr} & 
      \colhead{Gyr} & \colhead{mag} &
      \colhead{$10^{11} M_{\odot}$} & \colhead{$M_{\odot} \rm yr^{-1}$} & 
      \colhead{$(M/L_V)_{\odot}$} & \colhead{V} & \colhead{U-V}}
    \startdata 
1030-32 &
 2.34$_{- 0.12}^{+ 0.02}$ &
 1.01$_{- 0.21}^{+ 0.12}$ &
 0.040$_{- 0.030}^{+ 0.080}$ &
  0.0$_{-  0.0}^{+  0.4}$ &
 1.40$_{- 0.19}^{+ 0.38}$ &
     0.0$_{-     0.0}^{+     0.1}$ &
 0.85$_{- 0.07}^{+ 0.28}$ &
  -23.29$_{-    0.03}^{+    0.17}$ &
 0.70$_{- 0.07}^{+ 0.11}$ \\
      1030-101&
 1.77$_{- 0.11}^{+ 0.09}$ &
 0.57$_{- 0.17}^{+ 0.71}$ &
 0.020$_{- 0.010}^{+ 0.130}$ &
  1.5$_{-  0.8}^{+  0.5}$ &
 3.16$_{- 1.14}^{+ 0.75}$ &
     0.0$_{-     0.0}^{+     4.4}$ &
 2.42$_{- 0.54}^{+ 0.63}$ &
  -23.03$_{-    0.28}^{+    0.14}$ &
 1.18$_{- 0.11}^{+ 0.10}$ \\
      1030-301&
1.893   &
 0.20$_{- 0.10}^{+ 0.60}$ &
 0.040$_{- 0.030}^{+ 0.210}$ &
  2.8$_{-  0.5}^{+  0.2}$ &
 4.48$_{- 0.93}^{+ 1.24}$ &
    87.0$_{-    85.3}^{+   294.8}$ &
 3.25$_{- 0.60}^{+ 0.85}$ &
  -23.10$_{-    0.03}^{+    0.05}$ &
 1.10$_{- 0.06}^{+ 0.14}$ \\
      1030-609&
1.800   &
 0.20$_{- 0.10}^{+ 0.52}$ &
 0.120$_{- 0.095}^{+ 9.880}$ &
  1.8$_{-  0.2}^{+  0.2}$ &
 0.70$_{- 0.17}^{+ 0.29}$ &
   159.8$_{-   111.1}^{+   147.9}$ &
 0.91$_{- 0.20}^{+ 0.35}$ &
  -22.46$_{-    0.02}^{+    0.05}$ &
 0.35$_{- 0.05}^{+ 0.07}$ \\
      1030-807&
2.367   &
 0.72$_{- 0.52}^{+ 0.09}$ &
 0.100$_{- 0.085}^{+ 0.020}$ &
  0.0$_{-  0.0}^{+  1.4}$ &
 0.97$_{- 0.01}^{+ 0.55}$ &
     1.0$_{-     0.9}^{+     6.3}$ &
 0.59$_{- 0.00}^{+ 0.37}$ &
  -23.29$_{-    0.01}^{+    0.04}$ &
 0.46$_{- 0.01}^{+ 0.11}$ \\
     1030-1531&
2.613   &
 0.57$_{- 0.17}^{+ 0.57}$ &
 0.650$_{- 0.350}^{+ 9.350}$ &
  0.8$_{-  0.1}^{+  0.1}$ &
 1.67$_{- 0.19}^{+ 0.62}$ &
   227.1$_{-    56.3}^{+    72.7}$ &
 0.56$_{- 0.06}^{+ 0.18}$ &
  -23.92$_{-    0.05}^{+    0.04}$ &
 0.06$_{- 0.02}^{+ 0.07}$ \\
     1030-1611&
 1.93$_{- 0.14}^{+ 0.14}$ &
 0.57$_{- 0.37}^{+ 1.04}$ &
 0.120$_{- 0.110}^{+ 0.180}$ &
  1.7$_{-  1.1}^{+  0.6}$ &
 2.25$_{- 0.67}^{+ 0.52}$ &
    20.7$_{-    20.7}^{+    68.6}$ &
 1.97$_{- 0.41}^{+ 0.98}$ &
  -22.85$_{-    0.17}^{+    0.36}$ &
 0.96$_{- 0.12}^{+ 0.15}$ \\
     1030-1813&
 2.56$_{- 0.02}^{+ 0.14}$ &
 0.51$_{- 0.00}^{+ 0.21}$ &
 0.020$_{- 0.010}^{+ 0.060}$ &
  0.8$_{-  0.5}^{+  0.1}$ &
 5.04$_{- 1.25}^{+ 0.93}$ &
     0.0$_{-     0.0}^{+     1.7}$ &
 0.96$_{- 0.15}^{+ 0.20}$ &
  -24.42$_{-    0.13}^{+    0.03}$ &
 0.72$_{- 0.09}^{+ 0.07}$ \\
     1030-1839&
2.312   &
 2.00$_{- 1.49}^{+ 0.75}$ &
 4.000$_{- 3.800}^{+ 6.000}$ &
  1.3$_{-  0.4}^{+  0.3}$ &
 2.83$_{- 1.23}^{+ 0.94}$ &
   143.2$_{-    89.7}^{+   129.0}$ &
 1.83$_{- 0.82}^{+ 0.49}$ &
  -23.22$_{-    0.08}^{+    0.07}$ &
 0.53$_{- 0.11}^{+ 0.09}$ \\
     1030-2026&
2.511   &
 1.14$_{- 0.57}^{+ 0.29}$ &
 0.200$_{- 0.080}^{+ 0.050}$ &
  0.4$_{-  0.1}^{+  0.9}$ &
 2.68$_{- 0.09}^{+ 1.45}$ &
     6.0$_{-     0.7}^{+    27.2}$ &
 1.11$_{- 0.04}^{+ 0.59}$ &
  -23.71$_{-    0.01}^{+    0.01}$ &
 0.78$_{- 0.05}^{+ 0.10}$ \\
     1030-2329&
2.236   &
 0.81$_{- 0.60}^{+ 0.33}$ &
 0.150$_{- 0.140}^{+ 0.050}$ &
  0.7$_{-  0.4}^{+  1.2}$ &
 1.44$_{- 0.32}^{+ 0.41}$ &
     5.8$_{-     5.8}^{+    34.9}$ &
 1.15$_{- 0.26}^{+ 0.37}$ &
  -23.00$_{-    0.02}^{+    0.05}$ &
 0.73$_{- 0.10}^{+ 0.06}$ \\
     1030-2559&
 2.39$_{- 0.16}^{+ 0.05}$ &
 0.57$_{- 0.17}^{+ 1.04}$ &
 0.010$_{- 0.000}^{+ 0.190}$ &
  0.7$_{-  0.7}^{+  0.5}$ &
 2.10$_{- 0.67}^{+ 0.68}$ &
     0.0$_{-     0.0}^{+     2.5}$ &
 1.07$_{- 0.23}^{+ 0.62}$ &
  -23.32$_{-    0.20}^{+    0.12}$ &
 0.83$_{- 0.11}^{+ 0.15}$ \\
     1030-2728&
2.504   &
 0.51$_{- 0.22}^{+ 0.06}$ &
 0.080$_{- 0.070}^{+ 0.020}$ &
  0.9$_{-  0.3}^{+  0.4}$ &
 2.34$_{- 0.42}^{+ 0.28}$ &
     6.4$_{-     6.4}^{+     4.0}$ &
 1.04$_{- 0.18}^{+ 0.12}$ &
  -23.63$_{-    0.02}^{+    0.02}$ &
 0.69$_{- 0.06}^{+ 0.05}$ \\
     1030-2927&
 1.82$_{- 0.16}^{+ 0.20}$ &
 0.51$_{- 0.41}^{+ 0.77}$ &
 0.150$_{- 0.140}^{+ 0.150}$ &
  1.3$_{-  1.3}^{+  0.5}$ &
 1.08$_{- 0.38}^{+ 0.31}$ &
    31.7$_{-    31.7}^{+    50.1}$ &
 1.23$_{- 0.46}^{+ 0.35}$ &
  -22.61$_{-    0.32}^{+    0.30}$ &
 0.63$_{- 0.15}^{+ 0.12}$ \\
        1256-0&
 2.31$_{- 0.07}^{+ 0.05}$ &
 0.57$_{- 0.28}^{+ 0.44}$ &
 0.080$_{- 0.070}^{+ 0.070}$ &
  1.2$_{-  0.6}^{+  0.6}$ &
 4.06$_{- 0.78}^{+ 1.07}$ &
     5.2$_{-     5.2}^{+    34.2}$ &
 1.51$_{- 0.27}^{+ 0.46}$ &
  -23.82$_{-    0.08}^{+    0.11}$ &
 0.90$_{- 0.08}^{+ 0.10}$ \\
      1256-142&
 2.37$_{- 0.15}^{+ 0.05}$ &
 0.40$_{- 0.12}^{+ 2.35}$ &
 0.010$_{- 0.000}^{+ 0.390}$ &
  1.3$_{-  1.3}^{+  0.4}$ &
 3.22$_{- 1.08}^{+ 1.00}$ &
     0.0$_{-     0.0}^{+    16.1}$ &
 1.46$_{- 0.48}^{+ 0.76}$ &
  -23.60$_{-    0.07}^{+    0.21}$ &
 0.89$_{- 0.11}^{+ 0.15}$ \\
      1256-519&
1.857   &
 0.40$_{- 0.30}^{+ 2.60}$ &
 0.250$_{- 0.240}^{+ 9.750}$ &
  3.1$_{-  1.0}^{+  0.3}$ &
 4.27$_{- 1.63}^{+ 3.48}$ &
   523.4$_{-   521.9}^{+   817.5}$ &
 4.32$_{- 1.60}^{+ 3.63}$ &
  -22.74$_{-    0.04}^{+    0.05}$ &
 1.14$_{- 0.16}^{+ 0.19}$ \\
     1256-1207&
 1.84$_{- 0.04}^{+ 0.05}$ &
 0.40$_{- 0.12}^{+ 0.74}$ &
 0.025$_{- 0.015}^{+ 0.125}$ &
  1.5$_{-  1.0}^{+  0.3}$ &
 2.06$_{- 0.52}^{+ 0.34}$ &
     0.0$_{-     0.0}^{+     9.8}$ &
 1.70$_{- 0.41}^{+ 0.29}$ &
  -22.96$_{-    0.11}^{+    0.09}$ &
 0.96$_{- 0.08}^{+ 0.08}$ \\
     1256-1967&
 2.02$_{- 0.09}^{+ 0.07}$ &
 0.20$_{- 0.10}^{+ 0.00}$ &
 0.025$_{- 0.015}^{+ 0.015}$ &
  1.4$_{-  0.1}^{+  0.4}$ &
 2.48$_{- 0.33}^{+ 0.40}$ &
     3.7$_{-     3.7}^{+    48.1}$ &
 0.99$_{- 0.07}^{+ 0.09}$ &
  -23.75$_{-    0.12}^{+    0.16}$ &
 0.53$_{- 0.03}^{+ 0.05}$ \\
     HDFS1-259&
2.249   &
 0.81$_{- 0.52}^{+ 0.63}$ &
 5.000$_{- 4.800}^{+ 5.000}$ &
  1.5$_{-  0.2}^{+  0.2}$ &
 2.00$_{- 0.60}^{+ 0.48}$ &
   287.1$_{-   108.3}^{+   166.4}$ &
 1.14$_{- 0.32}^{+ 0.27}$ &
  -23.36$_{-    0.03}^{+    0.04}$ &
 0.37$_{- 0.10}^{+ 0.08}$ \\
    HDFS1-1849&
 2.31$_{- 0.08}^{+ 0.09}$ &
 0.40$_{- 0.20}^{+ 0.74}$ &
 0.080$_{- 0.070}^{+ 0.120}$ &
  1.6$_{-  1.0}^{+  0.5}$ &
 3.47$_{- 0.86}^{+ 1.00}$ &
    35.0$_{-    35.0}^{+    91.4}$ &
 1.59$_{- 0.47}^{+ 0.60}$ &
  -23.59$_{-    0.13}^{+    0.13}$ &
 0.84$_{- 0.16}^{+ 0.11}$ \\
     HDFS2-509&
2.918   &
 0.57$_{- 0.00}^{+ 0.00}$ &
 0.100$_{- 0.000}^{+ 0.000}$ &
  0.0$_{-  0.0}^{+  0.1}$ &
 3.54$_{- 0.00}^{+ 0.70}$ &
    15.0$_{-     0.0}^{+     3.0}$ &
 0.52$_{- 0.04}^{+ 0.00}$ &
  -25.02$_{-    0.00}^{+    0.10}$ &
 0.34$_{- 0.05}^{+ 0.00}$ \\
    HDFS2-1099&
 2.73$_{- 0.17}^{+ 0.08}$ &
 1.01$_{- 0.30}^{+ 0.00}$ &
 0.150$_{- 0.050}^{+ 0.000}$ &
  0.0$_{-  0.0}^{+  0.4}$ &
 2.83$_{- 0.44}^{+ 0.89}$ &
     2.8$_{-     1.5}^{+     4.9}$ &
 0.76$_{- 0.05}^{+ 0.22}$ &
  -24.18$_{-    0.10}^{+    0.18}$ &
 0.59$_{- 0.03}^{+ 0.10}$ \\
    HDFS2-2046&
 2.24$_{- 0.04}^{+ 0.09}$ &
 0.29$_{- 0.08}^{+ 0.28}$ &
 0.012$_{- 0.002}^{+ 0.068}$ &
  0.8$_{-  0.6}^{+  0.5}$ &
 1.41$_{- 0.25}^{+ 0.50}$ &
     0.0$_{-     0.0}^{+     6.5}$ &
 0.71$_{- 0.11}^{+ 0.21}$ &
  -23.49$_{-    0.14}^{+    0.07}$ &
 0.51$_{- 0.07}^{+ 0.04}$ \\
    ECDFS-4454&
2.351   &
 0.72$_{- 0.31}^{+ 0.09}$ &
 0.120$_{- 0.020}^{+ 0.030}$ &
  0.1$_{-  0.1}^{+  1.2}$ &
 1.20$_{- 0.16}^{+ 0.92}$ &
     3.2$_{-     0.4}^{+    69.7}$ &
 0.62$_{- 0.05}^{+ 0.50}$ &
  -23.46$_{-    0.03}^{+    0.06}$ &
 0.45$_{- 0.05}^{+ 0.14}$ \\
    ECDFS-4511&
2.122   &
 2.75$_{- 0.55}^{+ 0.25}$ &
10.000$_{- 7.000}^{+ 0.000}$ &
  1.1$_{-  0.1}^{+  0.0}$ &
 3.95$_{- 0.47}^{+ 0.48}$ &
   165.9$_{-    34.9}^{+     1.5}$ &
 1.80$_{- 0.20}^{+ 0.17}$ &
  -23.60$_{-    0.03}^{+    0.02}$ &
 0.46$_{- 0.02}^{+ 0.04}$ \\
    ECDFS-4713&
2.309   &
 0.51$_{- 0.00}^{+ 0.21}$ &
 0.150$_{- 0.030}^{+ 0.150}$ &
  0.7$_{-  0.2}^{+  0.3}$ &
 2.73$_{- 0.14}^{+ 0.93}$ &
    80.0$_{-    40.1}^{+   112.0}$ &
 0.71$_{- 0.07}^{+ 0.23}$ &
  -24.21$_{-    0.05}^{+    0.02}$ &
 0.35$_{- 0.00}^{+ 0.06}$ \\
    ECDFS-5856&
 2.56$_{- 0.04}^{+ 0.12}$ &
 0.40$_{- 0.12}^{+ 0.50}$ &
 0.030$_{- 0.020}^{+ 0.090}$ &
  1.0$_{-  0.8}^{+  0.4}$ &
 2.74$_{- 0.68}^{+ 0.77}$ &
     0.0$_{-     0.0}^{+     9.6}$ &
 1.06$_{- 0.26}^{+ 0.31}$ &
  -23.78$_{-    0.13}^{+    0.06}$ &
 0.73$_{- 0.11}^{+ 0.09}$ \\
    ECDFS-6842&
 2.40$_{- 0.10}^{+ 0.08}$ &
 0.57$_{- 0.28}^{+ 0.24}$ &
 0.150$_{- 0.100}^{+ 0.050}$ &
  1.5$_{-  0.5}^{+  0.3}$ &
 5.35$_{- 1.47}^{+ 0.90}$ &
   103.1$_{-    72.4}^{+   137.7}$ &
 1.66$_{- 0.45}^{+ 0.31}$ &
  -24.02$_{-    0.11}^{+    0.15}$ &
 0.81$_{- 0.13}^{+ 0.06}$ \\
    ECDFS-6956&
2.037   &
 0.05$_{- 0.00}^{+ 0.46}$ &
 0.010$_{- 0.000}^{+ 9.990}$ &
  1.4$_{-  0.1}^{+  0.2}$ &
 0.56$_{- 0.00}^{+ 0.62}$ &
    43.9$_{-     0.0}^{+   402.3}$ &
 0.41$_{- 0.00}^{+ 0.38}$ &
  -23.10$_{-    0.09}^{+    0.00}$ &
 0.08$_{- 0.00}^{+ 0.13}$ \\
    ECDFS-9822&
1.612   &
 0.10$_{- 0.00}^{+ 0.30}$ &
 0.020$_{- 0.005}^{+ 1.980}$ &
  1.9$_{-  0.1}^{+  0.3}$ &
 0.68$_{- 0.07}^{+ 0.29}$ &
    25.2$_{-    19.0}^{+   270.1}$ &
 0.97$_{- 0.03}^{+ 0.45}$ &
  -22.36$_{-    0.02}^{+    0.10}$ &
 0.50$_{- 0.05}^{+ 0.05}$ \\
   ECDFS-11490&
 2.34$_{- 0.02}^{+ 0.07}$ &
 0.51$_{- 0.10}^{+ 0.06}$ &
 0.065$_{- 0.055}^{+ 0.015}$ &
  0.1$_{-  0.1}^{+  0.2}$ &
 0.95$_{- 0.11}^{+ 0.08}$ &
     0.7$_{-     0.7}^{+     0.6}$ &
 0.52$_{- 0.06}^{+ 0.03}$ &
  -23.41$_{-    0.07}^{+    0.04}$ &
 0.37$_{- 0.05}^{+ 0.04}$ \\
   ECDFS-12514&
2.024   &
 0.40$_{- 0.20}^{+ 0.17}$ &
10.000$_{- 9.800}^{+ 0.000}$ &
  1.3$_{-  0.1}^{+  0.1}$ &
 1.25$_{- 0.28}^{+ 0.22}$ &
   369.1$_{-   105.7}^{+    91.7}$ &
 0.62$_{- 0.12}^{+ 0.09}$ &
  -23.51$_{-    0.03}^{+    0.07}$ &
 0.10$_{- 0.07}^{+ 0.05}$ \\
   ECDFS-13532&
2.336   &
 0.90$_{- 0.50}^{+ 0.53}$ &
 0.250$_{- 0.100}^{+ 0.750}$ &
  1.0$_{-  0.5}^{+  0.9}$ &
 1.88$_{- 0.46}^{+ 0.69}$ &
    26.8$_{-    18.4}^{+   217.2}$ &
 1.42$_{- 0.35}^{+ 0.59}$ &
  -23.05$_{-    0.02}^{+    0.05}$ &
 0.72$_{- 0.12}^{+ 0.08}$ \\
   ECDFS-16671&
 2.61$_{- 0.01}^{+ 0.02}$ &
 0.05$_{- 0.00}^{+ 0.00}$ &
 0.010$_{- 0.000}^{+ 0.030}$ &
  0.9$_{-  0.0}^{+  0.4}$ &
 1.08$_{- 0.05}^{+ 0.11}$ &
    83.9$_{-     1.0}^{+  1081.2}$ &
 0.26$_{- 0.01}^{+ 0.01}$ &
  -24.30$_{-    0.05}^{+    0.01}$ &
-0.16$_{- 0.05}^{+ 0.01}$ \\
     CDFS-6202&
2.225   &
 0.20$_{- 0.15}^{+ 0.08}$ &
 0.800$_{- 0.790}^{+ 9.200}$ &
  1.7$_{-  0.2}^{+  0.1}$ &
 1.26$_{- 0.41}^{+ 0.16}$ &
   649.8$_{-   583.6}^{+   145.1}$ &
 0.64$_{- 0.19}^{+ 0.07}$ &
  -23.48$_{-    0.03}^{+    0.04}$ &
 0.16$_{- 0.04}^{+ 0.03}$ \\

    \enddata
    \tablenotetext{a}{The continuum redshifts are those for which we
      give the confidence intervals.}  \tablecomments{The stellar
      population properties are derived by fitting the low resolution
      spectra together with the optical photometry by \cite{bc03}
      stellar population models. For the emission-line galaxies the
      redshift was fixed to \zs. The errors represent the 68\%
      confidence intervals derived using 200 Monte Carlo
      simulations. }
  \end{deluxetable*}
  \renewcommand\baselinestretch{1.}
}

\def\tabc{
  \renewcommand\baselinestretch{1.4}
  \begin{deluxetable*}{lllrlllllr}%[!t]
    \tabletypesize{\scriptsize} 
    \tablecaption{Photometric modeling results\label{tab_phot}}
    \tablewidth{0pt}
    \tablehead{ \colhead{} & \colhead{} & \colhead{age} & 
      \colhead{$\tau$} & \colhead{$A_V$} & \colhead{$M$} & \colhead{SFR} & 
      \colhead{$M/L_V$} & \colhead{} & \colhead{} \\
      \colhead{ID} & \colhead{$z_{\rm phot}$\tablenotemark{a}} & 
      \colhead{Gyr} & \colhead{Gyr} & \colhead{mag} &
      \colhead{$10^{11} M_{\odot}$} & \colhead{$M_{\odot} \rm yr^{-1}$} & 
      \colhead{$(M/L_V)_{\odot}$} & \colhead{V\tablenotemark{a}} & 
      \colhead{U-V\tablenotemark{a}}}
    \startdata 
       1030-32&
 2.26$_{- 0.28}^{+ 0.24}$ &
 2.20$_{- 1.80}^{+ 0.55}$ &
 0.250$_{- 0.240}^{+ 0.150}$ &
  0.0$_{-  0.0}^{+  2.1}$ &
 2.35$_{- 1.09}^{+ 1.81}$ &
     0.2$_{-     0.2}^{+    10.0}$ &
 1.49$_{- 1.01}^{+ 3.82}$ &
  -23.25$_{-    0.35}^{+    0.46}$ &
 0.91$_{- 0.28}^{+ 0.71}$ \\
      1030-101&
 2.18$_{- 0.52}^{+ 0.08}$ &
 0.90$_{- 0.62}^{+ 1.50}$ &
 0.010$_{- 0.000}^{+ 0.290}$ &
  0.2$_{-  0.2}^{+  1.7}$ &
 2.70$_{- 0.76}^{+ 3.51}$ &
     0.0$_{-     0.0}^{+    15.8}$ &
 0.91$_{- 0.36}^{+ 4.48}$ &
  -23.92$_{-    0.09}^{+    0.95}$ &
 0.72$_{- 0.15}^{+ 1.39}$ \\
      1030-301&
 2.94$_{- 0.68}^{+ 0.06}$ &
 1.01$_{- 0.30}^{+ 0.59}$ &
 0.120$_{- 0.110}^{+ 0.080}$ &
  0.0$_{-  0.0}^{+  0.9}$ &
 5.21$_{- 1.05}^{+ 4.39}$ &
     1.2$_{-     1.2}^{+    11.0}$ &
 0.71$_{- 0.18}^{+ 2.38}$ &
  -24.91$_{-    0.16}^{+    0.92}$ &
 0.53$_{- 0.00}^{+ 0.31}$ \\
      1030-609&
 2.22$_{- 0.92}^{+ 0.42}$ &
 0.51$_{- 0.50}^{+ 2.24}$ &
 0.300$_{- 0.290}^{+ 9.700}$ &
  1.4$_{-  1.4}^{+  1.3}$ &
 1.59$_{- 1.19}^{+ 2.26}$ &
   148.7$_{-   148.3}^{+  4621.8}$ &
 0.99$_{- 0.71}^{+15.05}$ &
  -23.26$_{-    0.49}^{+    2.10}$ &
 0.36$_{- 0.24}^{+ 0.27}$ \\
      1030-807&
 2.44$_{- 0.54}^{+ 0.38}$ &
 0.10$_{- 0.09}^{+ 2.50}$ &
 0.030$_{- 0.020}^{+ 9.970}$ &
  2.2$_{-  2.0}^{+  0.8}$ &
 1.90$_{- 1.23}^{+ 4.19}$ &
   262.5$_{-   262.5}^{+  6455.7}$ &
 1.02$_{- 0.76}^{+24.17}$ &
  -23.42$_{-    0.47}^{+    0.86}$ &
 0.42$_{- 0.05}^{+ 0.16}$ \\
     1030-1531&
 2.74$_{- 0.06}^{+ 0.10}$ &
 1.28$_{- 0.87}^{+ 0.92}$ &
 1.000$_{- 0.900}^{+ 9.000}$ &
  0.5$_{-  0.5}^{+  0.3}$ &
 2.64$_{- 1.57}^{+ 1.71}$ &
   131.9$_{-   107.9}^{+   149.1}$ &
 0.70$_{- 0.34}^{+ 0.27}$ &
  -24.19$_{-    0.17}^{+    0.12}$ &
 0.17$_{- 0.19}^{+ 0.17}$ \\
     1030-1611&
 2.32$_{- 0.38}^{+ 0.06}$ &
 1.01$_{- 0.81}^{+ 0.88}$ &
 0.150$_{- 0.140}^{+ 0.150}$ &
  0.3$_{-  0.3}^{+  1.6}$ &
 1.85$_{- 0.55}^{+ 1.21}$ &
     1.9$_{-     1.9}^{+    26.7}$ &
 0.93$_{- 0.44}^{+ 0.89}$ &
  -23.50$_{-    0.09}^{+    0.67}$ &
 0.46$_{- 0.00}^{+ 0.59}$ \\
     1030-1813&
 2.32$_{- 0.16}^{+ 0.16}$ &
 0.57$_{- 0.47}^{+ 1.43}$ &
 0.030$_{- 0.020}^{+ 9.970}$ &
  0.7$_{-  0.7}^{+  2.3}$ &
 3.29$_{- 1.59}^{+11.02}$ &
     0.0$_{-     0.0}^{+  2417.3}$ &
 0.91$_{- 0.00}^{+ 7.69}$ &
  -24.14$_{-    0.71}^{+    0.98}$ &
 0.17$_{- 0.00}^{+ 1.06}$ \\
     1030-1839&
 2.58$_{- 0.38}^{+ 0.20}$ &
 1.61$_{- 1.32}^{+ 0.79}$ &
 0.800$_{- 0.720}^{+ 9.200}$ &
  1.0$_{-  1.0}^{+  0.7}$ &
 3.78$_{- 2.59}^{+ 1.87}$ &
    96.2$_{-    90.5}^{+   391.7}$ &
 1.53$_{- 1.15}^{+ 2.69}$ &
  -23.73$_{-    0.31}^{+    0.60}$ &
 0.28$_{- 0.03}^{+ 0.31}$ \\
     1030-2026&
 2.68$_{- 0.30}^{+ 0.10}$ &
 1.28$_{- 0.77}^{+ 0.62}$ &
 0.200$_{- 0.100}^{+ 0.100}$ &
  0.0$_{-  0.0}^{+  1.6}$ &
 2.44$_{- 0.37}^{+ 3.49}$ &
     2.7$_{-     0.6}^{+    98.9}$ &
 0.79$_{- 0.36}^{+ 1.97}$ &
  -23.97$_{-    0.13}^{+    0.35}$ &
 0.68$_{- 0.09}^{+ 0.69}$ \\
     1030-2329&
 2.30$_{- 0.30}^{+ 0.48}$ &
 0.20$_{- 0.15}^{+ 2.55}$ &
 0.050$_{- 0.040}^{+ 9.950}$ &
  2.1$_{-  2.1}^{+  0.9}$ &
 2.25$_{- 1.13}^{+ 3.98}$ &
    96.9$_{-    96.9}^{+  2030.0}$ &
 1.48$_{- 1.10}^{+10.71}$ &
  -23.20$_{-    0.67}^{+    0.62}$ &
 0.63$_{- 0.16}^{+ 0.12}$ \\
     1030-2559&
 2.18$_{- 0.20}^{+ 0.34}$ &
 0.57$_{- 0.47}^{+ 2.18}$ &
 0.100$_{- 0.090}^{+ 0.400}$ &
  1.6$_{-  1.6}^{+  1.4}$ &
 2.72$_{- 1.52}^{+ 2.37}$ &
    11.5$_{-    11.5}^{+   468.0}$ &
 1.96$_{- 1.54}^{+ 3.73}$ &
  -23.10$_{-    0.46}^{+    0.39}$ &
 0.89$_{- 0.29}^{+ 0.37}$ \\
     1030-2728&
 2.60$_{- 0.32}^{+ 0.32}$ &
 0.20$_{- 0.15}^{+ 1.23}$ &
 0.025$_{- 0.015}^{+ 9.975}$ &
  1.7$_{-  1.3}^{+  1.3}$ &
 3.19$_{- 1.36}^{+ 5.46}$ &
     4.8$_{-     4.8}^{+  3551.5}$ &
 1.17$_{- 0.68}^{+ 2.56}$ &
  -23.84$_{-    0.39}^{+    0.48}$ &
 0.57$_{- 0.11}^{+ 0.14}$ \\
     1030-2927&
 2.26$_{- 0.52}^{+ 0.24}$ &
 0.29$_{- 0.24}^{+ 2.11}$ &
 0.080$_{- 0.070}^{+ 9.920}$ &
  1.6$_{-  1.6}^{+  1.0}$ &
 2.10$_{- 0.85}^{+ 2.94}$ &
    93.1$_{-    93.1}^{+  1673.3}$ &
 1.06$_{- 0.61}^{+ 6.61}$ &
  -23.49$_{-    0.34}^{+    0.82}$ &
 0.50$_{- 0.12}^{+ 0.14}$ \\
        1256-0&
 2.26$_{- 0.08}^{+ 0.04}$ &
 1.01$_{- 0.51}^{+ 0.12}$ &
 0.120$_{- 0.110}^{+ 0.030}$ &
  0.0$_{-  0.0}^{+  0.8}$ &
 1.90$_{- 0.16}^{+ 0.78}$ &
     0.4$_{-     0.4}^{+     2.2}$ &
 0.82$_{- 0.22}^{+ 0.41}$ &
  -23.67$_{-    0.17}^{+    0.13}$ &
 0.65$_{- 0.02}^{+ 0.01}$ \\
      1256-142&
 2.18$_{- 0.12}^{+ 0.14}$ &
 2.20$_{- 2.00}^{+ 0.55}$ &
 0.300$_{- 0.290}^{+ 0.100}$ &
  0.1$_{-  0.1}^{+  2.3}$ &
 2.76$_{- 1.33}^{+ 2.03}$ &
     0.8$_{-     0.8}^{+    86.7}$ &
 1.61$_{- 0.96}^{+ 2.78}$ &
  -23.33$_{-    0.20}^{+    0.34}$ &
 0.92$_{- 0.18}^{+ 0.45}$ \\
      1256-519&
 2.00$_{- 0.06}^{+ 0.26}$ &
 0.20$_{- 0.10}^{+ 0.94}$ &
 0.030$_{- 0.020}^{+ 0.220}$ &
  2.7$_{-  1.5}^{+  0.3}$ &
 4.94$_{- 1.49}^{+ 1.58}$ &
    23.6$_{-    23.6}^{+   549.0}$ &
 2.92$_{- 1.11}^{+ 1.57}$ &
  -23.32$_{-    0.45}^{+    0.14}$ &
 1.34$_{- 0.64}^{+ 0.31}$ \\
     1256-1207&
 2.10$_{- 0.44}^{+ 0.02}$ &
 0.51$_{- 0.22}^{+ 0.77}$ &
 0.030$_{- 0.020}^{+ 0.120}$ &
  1.0$_{-  1.0}^{+  0.8}$ &
 2.37$_{- 0.90}^{+ 0.98}$ &
     0.0$_{-     0.0}^{+    13.6}$ &
 1.22$_{- 0.46}^{+ 2.21}$ &
  -23.47$_{-    0.03}^{+    0.77}$ &
 1.00$_{- 0.09}^{+ 1.52}$ \\
     1256-1967&
 2.16$_{- 0.72}^{+ 0.06}$ &
 0.10$_{- 0.00}^{+ 0.62}$ &
 0.012$_{- 0.002}^{+ 0.088}$ &
  1.9$_{-  1.9}^{+  0.2}$ &
 3.51$_{- 1.86}^{+ 0.48}$ &
     7.4$_{-     7.4}^{+   164.4}$ &
 1.03$_{- 0.65}^{+ 0.46}$ &
  -24.08$_{-    0.09}^{+    1.34}$ &
 0.32$_{- 0.04}^{+ 0.10}$ \\
     HDFS1-259&
 2.24$_{- 0.18}^{+ 0.76}$ &
 0.72$_{- 0.71}^{+ 2.03}$ &
 0.650$_{- 0.640}^{+ 9.350}$ &
  1.4$_{-  1.4}^{+  1.3}$ &
 2.21$_{- 1.88}^{+ 3.48}$ &
   212.2$_{-   212.2}^{+  6029.4}$ &
 1.12$_{- 0.91}^{+ 6.10}$ &
  -23.49$_{-    0.95}^{+    0.31}$ &
 0.35$_{- 0.23}^{+ 0.07}$ \\
    HDFS1-1849&
 2.24$_{- 0.14}^{+ 0.24}$ &
 0.20$_{- 0.15}^{+ 1.50}$ &
 0.040$_{- 0.030}^{+ 9.960}$ &
  2.1$_{-  1.9}^{+  0.9}$ &
 3.22$_{- 1.38}^{+ 3.65}$ &
    62.6$_{-    62.6}^{+  2610.0}$ &
 1.59$_{- 0.83}^{+ 2.75}$ &
  -23.52$_{-    0.37}^{+    0.23}$ &
 0.71$_{- 0.16}^{+ 0.07}$ \\
     HDFS2-509&
 2.78$_{- 0.04}^{+ 0.40}$ &
 0.57$_{- 0.37}^{+ 1.63}$ &
 0.120$_{- 0.055}^{+ 9.880}$ &
  0.5$_{-  0.5}^{+  1.5}$ &
 5.62$_{- 2.56}^{+17.93}$ &
    51.7$_{-    43.1}^{+  3051.7}$ &
 0.70$_{- 0.46}^{+ 0.80}$ &
  -25.02$_{-    0.28}^{+    0.20}$ &
 0.30$_{- 0.27}^{+ 0.05}$ \\
    HDFS2-1099&
 2.40$_{- 0.12}^{+ 0.10}$ &
 0.81$_{- 0.40}^{+ 0.21}$ &
 0.120$_{- 0.070}^{+ 0.030}$ &
  0.4$_{-  0.4}^{+  0.9}$ &
 2.47$_{- 0.75}^{+ 1.25}$ &
     3.2$_{-     2.5}^{+    26.0}$ &
 0.87$_{- 0.48}^{+ 0.81}$ &
  -23.88$_{-    0.14}^{+    0.23}$ &
 0.67$_{- 0.11}^{+ 0.06}$ \\
    HDFS2-2046&
 2.36$_{- 0.10}^{+ 0.14}$ &
 0.20$_{- 0.10}^{+ 0.52}$ &
 0.010$_{- 0.000}^{+ 0.090}$ &
  1.2$_{-  1.2}^{+  0.8}$ &
 2.05$_{- 0.86}^{+ 0.89}$ &
     0.0$_{-     0.0}^{+   242.2}$ &
 0.88$_{- 0.46}^{+ 0.16}$ &
  -23.67$_{-    0.25}^{+    0.15}$ &
 0.33$_{- 0.06}^{+ 0.16}$ \\
    ECDFS-4454&
 2.50$_{- 0.08}^{+ 0.12}$ &
 0.72$_{- 0.00}^{+ 0.19}$ &
 0.120$_{- 0.000}^{+ 0.030}$ &
  0.2$_{-  0.2}^{+  0.5}$ &
 1.94$_{- 0.38}^{+ 1.28}$ &
     5.2$_{-     2.4}^{+    16.6}$ &
 0.68$_{- 0.25}^{+ 2.36}$ &
  -23.89$_{-    0.19}^{+    0.32}$ &
 0.83$_{- 0.53}^{+ 0.69}$ \\
    ECDFS-4511&
 2.70$_{- 0.06}^{+ 0.12}$ &
 1.14$_{- 0.23}^{+ 0.14}$ &
 0.250$_{- 0.050}^{+ 0.050}$ &
  0.1$_{-  0.1}^{+  0.1}$ &
 4.67$_{- 1.19}^{+ 0.80}$ &
    26.0$_{-     7.3}^{+     8.5}$ &
 0.82$_{- 0.25}^{+ 0.44}$ &
  -24.64$_{-    0.13}^{+    0.24}$ &
 0.52$_{- 0.17}^{+ 0.20}$ \\
    ECDFS-4713&
 2.30$_{- 0.04}^{+ 0.36}$ &
 1.01$_{- 0.73}^{+ 1.74}$ &
 0.500$_{- 0.420}^{+ 9.500}$ &
  1.0$_{-  1.0}^{+  0.5}$ &
 5.21$_{- 3.81}^{+ 7.28}$ &
   203.0$_{-   194.0}^{+   508.4}$ &
 1.26$_{- 0.92}^{+ 0.89}$ &
  -24.29$_{-    0.48}^{+    0.14}$ &
 0.32$_{- 0.24}^{+ 0.04}$ \\
    ECDFS-5856&
 3.32$_{- 0.12}^{+ 0.40}$ &
 1.61$_{- 0.80}^{+ 0.09}$ &
 0.300$_{- 0.180}^{+ 0.200}$ &
  0.7$_{-  0.7}^{+  0.8}$ &
12.20$_{- 8.30}^{+10.70}$ &
    25.6$_{-    21.6}^{+   177.8}$ &
 2.75$_{- 2.25}^{+ 0.23}$ &
  -24.37$_{-    0.57}^{+    0.03}$ &
-0.12$_{- 0.16}^{+ 0.72}$ \\
    ECDFS-6842&
 2.96$_{- 0.64}^{+ 0.04}$ &
 0.90$_{- 0.19}^{+ 1.10}$ &
 0.150$_{- 0.030}^{+ 0.650}$ &
  0.2$_{-  0.2}^{+  1.3}$ &
 4.55$_{- 1.42}^{+16.54}$ &
     9.5$_{-     4.8}^{+   310.5}$ &
 0.77$_{- 0.30}^{+ 4.62}$ &
  -24.68$_{-    0.12}^{+    0.76}$ &
 0.40$_{- 0.13}^{+ 0.46}$ \\
    ECDFS-6956&
 2.86$_{- 0.12}^{+ 0.04}$ &
 0.90$_{- 0.10}^{+ 1.10}$ &
 0.200$_{- 0.000}^{+ 0.300}$ &
  0.0$_{-  0.0}^{+  0.2}$ &
 3.14$_{- 1.17}^{+ 3.46}$ &
    22.3$_{-     0.0}^{+    20.9}$ &
 0.63$_{- 0.16}^{+ 0.85}$ &
  -24.49$_{-    0.12}^{+    0.44}$ &
 0.36$_{- 0.42}^{+ 0.78}$ \\
    ECDFS-9822&
 2.60$_{- 0.02}^{+ 0.18}$ &
 0.81$_{- 0.09}^{+ 1.59}$ &
 0.150$_{- 0.030}^{+ 4.850}$ &
  0.3$_{-  0.3}^{+  1.3}$ &
 2.93$_{- 1.17}^{+ 8.27}$ &
    11.8$_{-     7.0}^{+   377.4}$ &
 0.76$_{- 0.55}^{+ 2.45}$ &
  -24.22$_{-    0.29}^{+    0.15}$ &
 0.98$_{- 0.59}^{+ 1.08}$ \\
   ECDFS-11490&
 3.58$_{- 0.68}^{+ 0.14}$ &
 0.81$_{- 0.24}^{+ 0.89}$ &
 0.150$_{- 0.050}^{+ 0.350}$ &
  0.0$_{-  0.0}^{+  0.8}$ &
 4.43$_{- 2.60}^{+15.61}$ &
    17.8$_{-     4.1}^{+   158.2}$ &
 0.61$_{- 0.41}^{+ 0.88}$ &
  -24.91$_{-    0.47}^{+    0.55}$ &
 0.08$_{- 0.34}^{+ 1.31}$ \\
   ECDFS-12514&
 2.40$_{- 0.28}^{+ 0.30}$ &
 0.51$_{- 0.50}^{+ 2.09}$ &
 0.120$_{- 0.110}^{+ 9.880}$ &
  0.2$_{-  0.2}^{+  1.4}$ &
 1.87$_{- 1.77}^{+ 8.11}$ &
    28.7$_{-    17.2}^{+  2246.3}$ &
 0.58$_{- 0.33}^{+ 2.81}$ &
  -24.02$_{-    0.38}^{+    0.69}$ &
-0.13$_{- 0.22}^{+ 0.30}$ \\
   ECDFS-13532&
 2.86$_{- 0.24}^{+ 0.16}$ &
 1.28$_{- 0.37}^{+ 0.92}$ &
 0.250$_{- 0.100}^{+ 0.250}$ &
  0.4$_{-  0.4}^{+  0.5}$ &
 3.83$_{- 1.99}^{+ 5.14}$ &
    12.3$_{-     8.4}^{+    34.4}$ &
 1.07$_{- 0.75}^{+ 4.76}$ &
  -24.14$_{-    0.20}^{+    0.45}$ &
 1.29$_{- 0.64}^{+ 0.73}$ \\
   ECDFS-16671&
 3.02$_{- 0.12}^{+ 0.04}$ &
 0.72$_{- 0.21}^{+ 1.28}$ &
 0.250$_{- 0.100}^{+ 1.750}$ &
  0.0$_{-  0.0}^{+  0.2}$ &
 2.68$_{- 0.65}^{+ 4.93}$ &
    81.9$_{-    11.4}^{+    76.4}$ &
 0.41$_{- 0.10}^{+ 0.49}$ &
  -24.79$_{-    0.17}^{+    0.14}$ &
-0.11$_{- 0.30}^{+ 0.37}$ \\
     CDFS-6202&
 3.04$_{- 0.64}^{+ 0.06}$ &
 0.72$_{- 0.00}^{+ 0.19}$ &
 0.150$_{- 0.000}^{+ 0.050}$ &
  0.0$_{-  0.0}^{+  0.2}$ &
 3.19$_{- 0.00}^{+ 0.68}$ &
    22.9$_{-     0.0}^{+    22.1}$ &
 1.45$_{- 1.02}^{+ 0.00}$ &
  -23.60$_{-    0.06}^{+    0.03}$ &
 0.36$_{- 0.08}^{+ 0.08}$ \\

    \enddata
    \tablenotetext{a}{Derived using method by \cite{ru01,ru03}}
    \tablecomments{The stellar population properties are derived by
      fitting the photometry by \cite{bc03} models, fixing the
      redshift to \zp. The \zp\ confidence intervals are derived using
      100 Monte Carlo simulations \citep{ru01,ru03}. We use the same
      simulations in combination with the best-fit \zp\ for each
      simulation to derive the 1$\sigma$ (68\%) confidence
      intervals of the stellar population properties.}
  \end{deluxetable*}
  \renewcommand\baselinestretch{1.}
}

\def\tabd{
  \renewcommand\baselinestretch{1.2}
  \begin{deluxetable}{lllrrcc}
    \centering
    \tabletypesize{\scriptsize} 
    \tablecaption{Comparison between photometric and spectroscopic
      redshifts ($\Delta z/(1+z)$)\label{tab_dz}}
    \tablewidth{0pt}
    \tablehead{ & \multicolumn{2}{c}{All} &  
      \multicolumn{2}{c}{Deep} & 
      \multicolumn{2}{c}{Wide} \\
      & \colhead{scat} & \colhead{offs} &
      \colhead{scat} & \colhead{offs} &
      \colhead{scat} & \colhead{offs}}
    \startdata
                           All  &    0.13  &    0.08  &    0.08  &    0.03  &    0.24  &    0.17 \\
                $z_{\rm cont}$  &    0.13  &    0.06  &    0.10  &    0.01  &    0.31  &    0.21 \\
               $z_{\rm lines}$  &    0.10  &    0.07  &    0.06  &    0.03  &    0.24  &    0.16 \\
                          DRGs  &    0.11  &    0.08  &    0.07  &   -0.00  &    0.32  &    0.22 \\
          DRGs, $z_{\rm cont}$  &    0.15  &    0.01  &    0.09  &   -0.05  &    0.35  &    0.23 \\
         DRGs, $z_{\rm lines}$  &    0.07  &    0.04  &    0.06  &    0.03  &    0.30  &    0.21 \\
                      non-DRGs  &    0.13  &    0.08  &    0.13  &    0.08  &    0.18  &    0.12 \\
      non-DRGs, $z_{\rm cont}$  &    0.13  &    0.08  &    0.12  &    0.08  &       -  &       - \\
     non-DRGs, $z_{\rm lines}$  &    0.14  &    0.09  &    0.16  &    0.11  &    0.09  &    0.06 \\

    \enddata
    \tablecomments{In this table we give the scatter and systematic
      offset (\dzs) between the photometric and spectroscopic
      redshifts for the full sample, and different subsamples.}
  \end{deluxetable}
  \renewcommand\baselinestretch{1.0} 
}

\def\tabe{
  \begin{deluxetable}{lrrlrlrl}
    \tabletypesize{\scriptsize} 
    \tablecaption{Comparison between photometrically and spectroscopically 
      derived properties\label{tab_scatter}}
    \tablewidth{0pt} 
    \tablehead{ \multicolumn{2}{l}{Quantity} &  
      \multicolumn{2}{c}{$Q_{\rm ph} (z_{\rm ph})$} & 
      \multicolumn{2}{c}{$Q_{\rm ph} (z_{\rm ph})$} &
      \multicolumn{2}{c}{$Q_{\rm ph} (z_{\rm sp})$} \\
      \multicolumn{2}{l}{} & \multicolumn{2}{c}{vs.} & 
      \multicolumn{2}{c}{vs.} & \multicolumn{2}{c}{vs.} \\
      \multicolumn{2}{l}{} &
      \multicolumn{2}{c}{$Q_{\rm sp} (z_{\rm sp})$} & 
      \multicolumn{2}{c}{$Q_{\rm ph} (z_{\rm sp})$} &
      \multicolumn{2}{c}{$Q_{\rm sp} (z_{\rm sp})$} \\
      \colhead{} & \colhead{} & 
      \multicolumn{1}{c}{offs} & \multicolumn{1}{c}{scat} &
      \multicolumn{1}{c}{offs} & \multicolumn{1}{c}{scat} &
      \multicolumn{1}{c}{offs} & \multicolumn{1}{c}{scat}
    }
    \startdata
                 age &       all &    1.22 &    2.70 &    1.00 &    1.76 &    1.06 &    2.67 \\
                     &      deep &    1.04 &    2.55 &    1.02 &    1.69 &    1.01 &    2.73 \\
             $A_{V}$ &       all &   -0.34 &    1.03 &   -0.30 &    0.59 &    0.05 &    0.47 \\
                     &      deep &   -0.12 &    0.88 &   -0.11 &    0.35 &   -0.02 &    0.45 \\
              $\tau$ &       all &    0.89 &    4.18 &    0.93 &    3.00 &    1.00 &    4.77 \\
                     &      deep &    0.80 &    3.13 &    0.82 &    2.11 &    0.76 &    5.74 \\
                mass &       all &    1.38 &    2.23 &    1.11 &    1.78 &    1.14 &    1.89 \\
                     &      deep &    1.28 &    2.05 &    1.11 &    1.72 &    1.06 &    1.78 \\
                 SFR &       all &    0.81 &    4.13 &    0.83 &    4.13 &    0.98 &    2.63 \\
                     &      deep &    0.88 &    3.51 &    0.94 &    3.49 &    0.78 &    3.00 \\
           $M/L_{V}$ &       all &    1.05 &    2.06 &    0.97 &    1.78 &    1.09 &    1.75 \\
                     &      deep &    1.00 &    1.93 &    0.98 &    1.76 &    1.05 &    1.77 \\
                 $V$ &       all &   -0.38 &    0.60 &   -0.34 &    0.56 &   -0.01 &    0.07 \\
                     &      deep &   -0.21 &    0.40 &   -0.19 &    0.38 &   -0.00 &    0.04 \\
             $(U-V)$ &       all &   -0.06 &    0.30 &   -0.06 &    0.21 &    0.02 &    0.08 \\
                     &      deep &   -0.08 &    0.20 &   -0.06 &    0.14 &    0.01 &    0.06 \\

    \enddata
    \tablecomments{In this table we list for all properties the
    scatter and systematic offset between the photometrically and
    spectroscopically derived properties ($Q_{\rm ph}$(\zp)
    vs. $Q_{\rm sp}$(\zs)). Furthermore, we examine whether this
    scatter between the photometric and spectroscopic properties is
    mainly driven by the precise redshift measurement ($Q_{\rm
    ph}$(\zp) vs. $Q_{\rm ph}$(\zs)), or the higher resolution 
    spectral shape ($Q_{\rm ph}$(\zs) vs. $Q_{\rm sp}$(\zs))}
  \end{deluxetable}
}

\documentclass[aaspp4,11pt,preprint,apjfonts]{emulateapj}

\slugcomment{Accepted for publication in ApJ}
\shorttitle{A NIR Spectroscopic Survey of $z\sim2.3$ Galaxies}
\shortauthors{Kriek et al.}

\newcommand{\zs}{$z_{\rm spec}$}
\newcommand{\zp}{$z_{\rm phot}$}

\newcommand{\dz}{$(z_{\rm phot} - z_{\rm spec}) / (1+z_{\rm spec})$}
\newcommand{\dzs}{$\Delta z/(1+z)$}

\begin{document}

\title{A Near-Infrared Spectroscopic Survey of K-selected Galaxies at
  $\lowercase{z}\sim2.3$: Redshifts and Implications for Broadband
  Photometric Studies \altaffilmark{1,2}}

\author{Mariska Kriek\altaffilmark{3,4,5}, Pieter G. van
  Dokkum\altaffilmark{4}, Marijn Franx\altaffilmark{3}, Garth D.
  Illingworth\altaffilmark{6}, Danilo Marchesini\altaffilmark{4}, Ryan
  Quadri\altaffilmark{4,3}, Gregory Rudnick\altaffilmark{7}, Edward
  N. Taylor\altaffilmark{3}, Natascha M. F\"orster
  Schreiber\altaffilmark{8}, Eric Gawiser\altaffilmark{4,9}, Ivo
  Labb{\'e}\altaffilmark{10}, Paulina Lira\altaffilmark{11} and Stijn
  Wuyts\altaffilmark{3,12}}

\email{mariska@astro.princeton.edu}

\altaffiltext{1}{Based on observations obtained at the Gemini
  Observatory, which is operated by the Association of Universities for
  Research in Astronomy, Inc., under a cooperative agreement with the
  NSF on behalf of the Gemini partnership.}

\altaffiltext{2}{Fully reduced spectra and corresponding data products
  will be made public at http://www.astro.yale.edu/MUSYC on January
  15, 2008}

\altaffiltext{3}{Leiden Observatory, Leiden University,
  PO Box 9513, 2300 RA Leiden, Netherlands}

\altaffiltext{4}{Department of Astronomy, Yale University, New Haven,
  CT 06520}

\altaffiltext{5}{{\em Current address:} Department of Astrophysical
  Sciences, Princeton University, Princeton, NJ 08544}

\altaffiltext{6}{UCO/Lick Observatory, University of California, Santa
  Cruz, CA 95064}

\altaffiltext{7}{Goldberg Fellow, National Optical Astronomy Observatory, 
  950 North Cherry Avenue, Tucson, AZ 85719}

\altaffiltext{8}{Max-Planck-Institut f\"ur extraterrestrische Physik,
  Giessenbachstrasse, Postfach 1312, D-85748 Garching, Germany}

\altaffiltext{9}{Department of Physics and Astronomy, Rutgers University, 
  Piscataway, NJ 08854}

\altaffiltext{10}{Hubble Fellow, Carnegie Observatories, 813 Santa
  Barbara Street, Pasadena, CA 91101}

\altaffiltext{11}{Departamento de Astronom{\'i}a, Universidad de Chile, 
  Casilla 36-D, Santiago, Chile}

\altaffiltext{12}{Harvard-Smithsonian Center for Astrophysics, 60
  Garden Street, Cambridge, MA 02138}

\begin{abstract} 
  Using the Gemini Near-InfraRed Spectrograph (GNIRS), we have
  completed a near-infrared spectroscopic survey for $K$-bright
  galaxies at $z\sim2.3$, selected from the MUSYC survey. We derived
  spectroscopic redshifts from emission lines or from continuum
  features and shapes for all 36 observed galaxies. The continuum
  redshifts are driven by the Balmer/4000\,\AA\ break, and have an
  uncertainty in $\Delta{}z/(1+z)$ of $<0.019$. We use this unique
  sample to determine, for the first time, how accurately redshifts
  and other properties of massive high-redshift galaxies can be
  determined from broadband photometric data alone. We find that the
  photometric redshifts of the galaxies in our sample have a
  systematic error of 0.08 and a random error of 0.13 in
  $\Delta{}z/(1+z)$. The systematic error can be reduced by using
  optimal templates and deep photometry; the random error, however,
  will be hard to reduce below 5\%. The spectra lead to significantly
  improved constraints for stellar population parameters. For most
  quantities this improvement is about equally driven by the higher
  spectral resolution and by the much reduced redshift
  uncertainty. Properties such as the age, $A_V$, current star
  formation rate, and the star formation history are generally very
  poorly constrained with broadband data alone. Interestingly stellar
  masses and mass-to-light ratios are among the most stable parameters
  from broadband data. Nevertheless, photometric studies may
  overestimate the number of massive galaxies at $2<z<3$, and thus
  underestimate the evolution of the stellar mass density. Finally,
  the spectroscopy supports our previous finding that red galaxies
  dominate the high-mass end of the galaxy population at $z=2-3$.
\end{abstract}


\taba

\section{INTRODUCTION}

The extensive use of photometric redshifts has greatly enhanced our
knowledge of the $z=2-3$ universe. While color criteria, such as the
Lyman break technique \citep[][]{st96a,st96b}, distant red galaxy
selection \citep[DRGs,][]{fr03,vd03} and BzK selection \citep{da04}
provide an easy identification of high-redshift galaxies, photometric
redshifts allow the study of apparently unbiased samples. Our current
understanding of the evolution of the mass-density
\citep[e.g.,][]{ru01,ru03,ru06,di03,dr05} and the luminosity function
\citep[e.g.,][]{dah05,sa06,ma07}, the nature of massive high-redshift
galaxies \citep[e.g.,][]{fo04,la05,vd06,pa06}, and galaxy clustering
\citep[e.g.,][]{da03,qu07a,fo07} essentially all rely on photometric
redshifts.

Ideally, all these studies would have been based on spectroscopic
redshifts. However, obtaining spectroscopic redshifts for the required
samples is hampered by several obstacles. Due to their faintness,
obtaining redshifts of high-redshift galaxies requires long
integrations on 8-10m class telescopes. The largest samples of
spectroscopically confirmed high redshift galaxies number in the 1000s
\citep[e.g.,][]{st03}, several orders of magnitude short of the
largest photometric samples. Furthermore, and more fundamentally,
current spectroscopic samples are strongly biased towards blue, star
forming galaxies which are bright in the observer's optical (Lyman
break galaxies, or LBGs). However, it has become clear that the
typical massive galaxy at this redshift range is red in the rest-frame
optical and faint in the rest-frame ultra-violet (UV), and blue LBGs
constitute only $\sim$20\% of massive galaxies at $z=2-3$
\citep{vd06}.

Obtaining spectroscopic redshifts for typical massive galaxies
requires deep spectroscopy at near-infrared (NIR)
wavelengths. Unfortunately, NIR observations are complicated by the
combination of the high sky brightness, numerous bright and variable
night sky lines and strong atmospheric absorption bands, and the
limited field of view of current and planned NIR spectrographs. Thus,
obtaining spectroscopic redshifts for hundreds or thousands of
galaxies with $K\sim21$ (the typical brightness of galaxies with
$M>10^{11}\,M_{\odot}$ at $z\sim2.5$) will not be feasible in the
foreseeable future. Until the next generation of space missions and
$>20$\,m ground-based telescopes we remain largely dependent on
photometric redshifts for studies of large and faint galaxy samples
beyond $z>1.5$.

Our provisional dependency on broadband photometric studies requires a
more accurate calibration and understanding of the involved
systematics. The current spectroscopic samples used for calibration of
photometric high-redshift studies are based primarily on optical
spectroscopy. As these samples are biased towards un-obscured
star-forming galaxies, their calibration may not be representative for
the total sample of massive galaxies. Photometric properties of red,
massive galaxies at high redshift are poorly calibrated, and since red
galaxies dominate the high mass end at $2<z<3$, systematics may have
large effects on the final results.

NIR spectroscopy on a substantial, unbiased sample of massive,
high-redshift galaxies is needed to test our photometric studies, and
obtain insights regarding possible systematic effects. The Gemini
Near-InfraRed Spectrograph \citep[GNIRS,][]{el06} is especially
well-suited for spectroscopy of $z\sim2.3$ galaxies, due to the large
wavelength coverage offered by the cross-disperser (0.9-2.5
$\mu$m). This coverage offers two advantages: there is a large
probability of finding emission lines, and it allows the
characterization of the continuum emission, which is particularly
important for galaxies without detected lines. In \cite{kr06b} we
found that a substantial fraction of the massive galaxies at
$z\sim2.3$ have no detected emission lines. For these galaxies we
derived spectroscopic redshifts by modeling the stellar continuum,
driven by the Balmer/4000 \AA\ break \citep{kr06a}.

In this paper we present our full survey of $K$-selected galaxies at
$z\sim2.3$, conducted with GNIRS on Gemini-South between September
2004 and March 2007. In total we integrated more than $\sim$80 hours,
divided over 6 observing runs, on a sample of 36 galaxies. In previous
papers based on preliminary results of this survey we discussed the
stellar populations \citep{kr06b} and the origin of the line emission
\citep{kr07}. In this paper we will give an overview of the total
survey (\S~\ref{dat}), compare photometric and spectroscopic redshifts
(\S~\ref{red}) and stellar populations properties (\S~\ref{imp}), and
discuss the implications for photometric studies (\S~\ref{imp}).

Throughout the paper we assume a $\Lambda$CDM cosmology with
$\Omega_{\rm m}=0.3$, $\Omega_{\rm \Lambda}=0.7$, and $H_{\rm
0}=70$~km s$^{-1}$ Mpc$^{-1}$.  The broadband magnitudes are given in
the Vega-based photometric system unless stated
otherwise. Furthermore, we will measure the scatter and the offset
between various properties using the normalized biweight mean absolute
deviation and the biweight mean, respectively \citep{be90}. As
biweight statistics are less sensitive towards outliers than the
normal mean, and more efficient than the median, they are most
appropriate for the small sample sizes in this work.

\figa %Figure1: J-K and R-K versus z_phot

\section{DATA}\label{dat}

\subsection{Sample Selection}

The galaxies studied in this paper are selected from the
MUlti-wavelength Survey by Yale-Chile
\citep[MUSYC,][]{ga06,qu07b}. This survey consist of optical imaging
(UBVRIz) of four 30\arcmin\ $\times$ 30\arcmin\ fields, shallow NIR
imaging (JHK) over the same area, and deeper NIR imaging over four
10\arcmin\ $\times$ 10\arcmin\ fields. The depth of the deep and wide
NIR photometry is $K\sim21$ and $K\sim20$ (5$\sigma$)
respectively. The spectroscopic follow-up presented in this paper is
selected from the deep fields HDF-South, 1030, and 1256 \citep{qu07b},
and the shallow extended Chandra Deep Field South (ECDFS, E.N. Taylor
et al. 2007, in preparation). One galaxy is selected from the Great
Observatories Origins Deep Survey \citep[GOODS;][]{gi04}. The
optical-to-NIR photometry that we used as part of this work is
described by S. Wuyts et al. (2007, in preparation).

We selected galaxies with $2.0\lesssim z_{\rm phot}\lesssim 2.7$ (see
\S~\ref{zphot}) and $K\lesssim 19.7$. This redshift interval is chosen as the
bright emission lines H$\beta$, [O\,{\sc iii}], H$\alpha$, and
[N\,{\sc ii}] fall in the $H$ and $K$ atmospheric windows. A few
galaxies had $2.7\lesssim z_{\rm phot}\lesssim 3.0$ at the time of
selection, but were observed because a large part of their confidence
intervals fell in the targeted redshift range, or because the galaxy
had a known \zs\ between 2.0 and 2.7
\citep[CDFS-6202,][]{vd05,dad05}. Due to catalog updates of the ECDFS
field, the final \zp\ of several more galaxies scattered out of the
selected redshift range. The fact that the ECDFS catalog was still in
a preliminary stage at the time of selection may complicate some of
the interpretation in this work. For this reason and the fact that the
ECDFS is the only field with shallow NIR photometry, the analysis in
this paper will also focus on the subsample excluding the ECDFS.

In total we obtained usable NIR spectra for a sample of 36
galaxies. For $\sim$4 additional galaxies we obtained empty spectra
due to mis-alignment or extremely bad weather conditions.

\figb %Figure 2: Histograms, comparison spec. and phot. sample

It is important to establish whether our sample is representative for
the galaxy population at $z\sim2.5$. In Figure~\ref{mass_vs_k_sel} we
compare the distributions of $J-K$, $R-K$, rest-frame $U-V$ color and
\zp\ of our spectroscopic sample with a photometric mass-limited
sample ($>10^{11} M_{\odot}$) at $2<z<3$. For the latter we use the
deep MUSYC fields, as the wide NIR data are too shallow to extract a
mass-limited sample \citep{vd06}. According to a Mann-Whitney and a
Kolmorov-Smirnov test, the photometric properties $J-K$, $R-K$,
rest-frame $U-V$ color (as derived from the photometry) and \zp\ of
the GNIRS sample are representative for a photometric mass-limited
sample at $2<z<3$ (see probabilities in panels of
Fig.~\ref{mass_vs_k_sel}). The Mann-Whitney test assesses whether the
two sample populations are consistent with the same mean of
distribution, while the K-S test examines whether the two samples
could have been drawn from the same parent distribution. For both
tests the probability should be greater than the 0.05 significance
level.

The galaxies targeted with GNIRS are all bright in $K$ ($<19.7$). In
order to examine if bright galaxies are a biased sub-sample of the
total mass-limited sample, we split the sample in bright and faint
members. Figure~\ref{mass_vs_k_sel} shows that the bright and faint
members have the same distribution of rest-frame $U-V$ and observed
$R-K$ colors. The main difference between the bright and faint members
is the redshift distribution: almost all $K$-bright galaxies have
\zp$<2.3$. This also causes their bluer $J-K$ colors: in contrast to
the $R$-band, the $J$-band does not fall entirely bluewards of the
optical break for $z<2.3$. The redshift dependence of $J-K$ is clearly
visible in Figure~\ref{rep}. However, except for the difference in
redshift and presumably stellar mass, we see no hints that the bright
and faint members of a mass-limited sample at $2<z<3$ have different
stellar populations. Thus, although the spectroscopic sample may have
similar stellar population properties as $K$-bright galaxies, it is
less representative for a $K$-bright sample at $2<z<3$, as the median
redshift and its corresponding distribution is substantially
different.

\figc %Figure 3: Compare observing sequences
\figd %Figure 4: Compare spectroscopic and photometric colors
\fige %Figure 5: Overview spectra
\addtocounter{figure}{-1}
\figf %Figure 5: Overview spectra

\subsection{NIR spectra}\label{nir_spectra}

We observed the full sample of 36 galaxies with GNIRS in
cross-dispersed mode, in combination with the 32 lines~mm$^{-1}$
grating and the 0\farcs675 slit. This configuration resulted in a
spectral resolution of $R\sim1000$. The galaxies were observed during
six observing runs in 2004 September (program GS-2004B-Q-38), 2005 May
(program GS-2005A-Q-20), 2006 January (program GS-2005B-C-12) and 2006
February (program GS-2006A-C-6), 2006 December (program GS-2006B-C-5)
and 2007 March (program GS-2007A-C-9). During the first two runs most
time was lost due to bad weather, and only a handful of galaxies was
observed under mediocre weather condition (seeing $\sim 1$\arcsec).
The weather was excellent throughout the full 3rd and 4th run, and we
reached a median seeing of $\sim$0\farcs5. The conditions were
slightly worse during the last two runs, with a median seeing of
$\sim$0\farcs7, and some time was lost due to clouds.

We observed the galaxies following an ABA\arcmin B\arcmin\ on-source
dither pattern, such that we can use the average of the previous and
following exposures as sky frame. This cancels sky variation and
reduces the noise in the final frame. All targets were acquired by
blind offsets from nearby stars. The individual exposures are 5
minutes for the galaxies observed during the first two runs, and 10
minutes for the remaining runs\footnote{An instrument upgrade after
  the first two runs improved the throughput and thus the quality of
  the spectra, and eliminated ``radiation events'' caused by
  radioactive coatings. This allowed longer integrations}. The total
integration times for all galaxies are listed in
Table~\ref{sample}. Before and after every observing sequence we
observe an A~V0 star, for the purpose of correcting for telluric
absorption. The final spectra of the two stars were combined to match
the target's airmass.

A detailed description of the reduction procedure of the GNIRS
cross-dispersed spectra is given in \cite{kr06a}. In summary, we
subtract the sky, mask cosmic rays and bad pixels, straighten the
spectra, combine the individual exposures, stitch the orders and
finally correct for the response function. 1D spectra are extracted by
summing all adjacent lines (along the spatial direction) with a mean
flux greater than 0.25 times the flux in the central row, using
optimal weighting. We also constructed ``low resolution'' binned
spectra from the 2D spectra for each galaxy following the method as
described in \cite{kr06a}. Each bin contains 80 ``good'' pixels (i.e.,
wavelength regions with high atmospheric transmission and low sky
emission), corresponding to 400 \AA\ per bin. Sky, transmission, and
noise spectra were constructed for each galaxy as well.

We assess the uncertainties on the low resolution spectra by splitting
the data in two sequences for several objects and comparing the
results. We find that the errors as derived from the photon noise
underestimate the true uncertainty for most galaxies. In
Figure~\ref{error} we show two examples. In order to obtain a better
consistency between the observing sequences, we increase the
uncertainties for all bins by quadratically adding 10\% of the average
flux in the spectrum.

We use the broadband NIR photometry to perform the absolute flux
calibration. For each galaxy we integrate the spectrum over the same
$J$, $H$ and $K$ filter curves as the photometry. We determine one
scaling factor per galaxy, from the difference between the
spectroscopic and photometric NIR magnitudes, and use this factor to
scale the NIR spectrum. We note that at this stage both the spectra
and the photometry may contain flux contributions by emission
lines. We extend our wavelength coverage by attaching the optical
photometry to the scaled spectra. For the emission-line galaxies we
subsequently remove the line fluxes from the affected bins, using the
best-fit to the emission lines as derived in \cite{kr07}.

As a quality check we compare the photometric NIR colors $J-H$, $J-K$
and $H-K$ to those derived from the spectra. The direct comparison for
the individual galaxies is presented in Figure~\ref{q_check}. As
expected, the scatter is larger for the galaxies with shallow NIR
photometry. The fraction of galaxies for which the spectroscopic and
photometric colors are consistent within 1$\sigma$ are listed in the
panels. For all colors these fractions are slightly lower than
expected, suggesting that the uncertainties may be slightly
underestimated. This may be partly due to color gradients in the
galaxies, as we use different apertures to determine the colors. The
spectroscopic apertures are rectangular, and depend on the extraction
radius and method, which differs per galaxy. On average the
spectroscopic apertures are 0\farcs675 by 1\farcs2, slightly smaller
than the circular photometric apertures which have a diameter of
$\sim$1\farcs4.

The final low resolution spectra and broadband photometry for the full
sample are presented in Figure~\ref{spec}. The galaxies are ordered by
total $K$ magnitude (see Table~\ref{sample}), starting with the
brightest galaxy. Remarkably, there is not a strong trend between
quality of the spectrum and the total $K$-band magnitude for the same
integration time. We suspect that the surface brightness of the object
plays an important role, as bright objects with low quality spectra
and typical integration times, such as ECDFS-4511 and ECDFS-9822, are
extended even in the MUSYC imaging, which have a image quality of
$\sim1$\arcsec.

\section{SPECTROSCOPIC VERSUS PHOTOMETRIC REDSHIFTS}
\label{red}

\subsection{Spectroscopic Redshifts and Galaxy Properties}
\label{zspec}

For 19 of the galaxies in the sample we detected one or more emission
lines, and thus for these galaxies we could determine exact
spectroscopic redshifts. The remaining galaxies may have no or very
faint emission lines, or the lines are expected in atmospheric
wavelength regions with low transmission or strong sky
emission. Fortunately, we can derive fairly precise redshifts from the
continuum emission alone, mainly due to the presence of the
Balmer/4000 \AA\ break in the NIR spectra \citep{kr06a}. For none of
the galaxies absorption lines are detected.

We fit the low resolution binned spectra together with the optical
photometry by \cite{bc03} stellar population models. We allow a grid
of 24 different ages (not allowing the galaxy to be older than the age
of the universe), and 31 different exponentially declining star
formation histories (SFHs) with the characteristic timescale ($\tau$)
varying between 10~Myr and 10~Gyr. We leave redshift as a free
parameter for the galaxies without emission lines. Furthermore, we
adopt the \cite{ca00} reddening law and allow 41 values for $A_V$
between 0 en 4 mag. We compute the $\chi^2$ surface as function of all
stellar population parameters. For all grid points we assume the
\cite{sa55} IMF, and solar metallicity. A \cite{ch03} or a \cite{kr01}
IMF yield stellar masses and SFRs which are a factor of $\sim2$
lower. The mass differences when using the stellar population library
by \cite{ma05} are discussed in \cite{kg07} and \cite{wu07a}.

We derive 1$\sigma$ confidence intervals on the redshifts and stellar
population properties using 200 Monte Carlo simulations. We vary all
bins of the low-resolution spectra according to their uncertainties,
and fit the simulated spectra using the same procedure as described
above. Next, we determine the contour in the original $\chi^2$ surface
that encompasses 68\% of the Monte Carlo simulations
\citep[see][]{pa03,kr06a}. The 1$\sigma$ confidence intervals for all
properties are the minimum and maximum values that are allowed within
this $\chi^2$ contour.

For the emission-line galaxies, we removed the emission line fluxes
from the spectra before fitting. This is different to our previous
method presented in \cite{kr06a} in which we mask the bins that are
contaminated by emission lines. The difference in modeling results
(although consistent within the errors) compared to \cite{kr06b,kr07}
are due to this improvement and updates of the broadband photometry
catalogs. All spectroscopic redshifts and corresponding stellar
population properties are listed in Table~\ref{tab_spec}.

\figg %Figure 7: z_cont vs z_lines
\figh %Figure 8: Systematics in continuum redshifts
\figi %Figure 6: templates
\tabb
\tabc

In order to test the accuracy of the continuum redshifts, we also fit
the emission-line galaxies with redshift as a free parameter. In
Figure~\ref{ztest} we compare the emission line redshifts with the
continuum redshifts. We find a scatter of \dzs$=0.019$ and no
significant systematic offset. In Figure~\ref{ztest2} we examine
causes of the errors. First, as the modeling is driven by the optical
break, we expect this method to be less accurate if the break falls
between atmospheric windows. In the Figure~\ref{ztest2}a we indeed
find that galaxies for which the break falls outside the spectrum or
in between the $J$ and $H$ band have less accurate $z_{\rm cont}$.

Furthermore, we expect the continuum redshifts to be more accurate for
galaxies with strong optical breaks. In the Figures~\ref{ztest2}b and
c we show the correlation with rest-frame $U-V$ color and observed
$R-K$ color. If we exclude the galaxies for which the break falls
between atmospheric windows ({\it gray symbols}), we indeed find that
blue galaxies with weak optical breaks, have less accurate continuum
redshifts. The continuum redshifts for galaxies without emission lines
might even be more accurate, as these galaxies generally have larger
breaks \citep{kr06b}.

Finally, in Figure~\ref{ztest2}d we examine the correlation between
$(z_{\rm cont}-z_{\rm line})/(z_{\rm line}+1)$ and the S/N per
bin. Remarkably, we do not find an obvious trend. However, the
different causes for errors and the small size of the sample may have
affected a possible correlation.

\figj % Figure 9: z_phot vs z_spec
\tabd

\subsection{Photometric Redshifts and Galaxy Properties}\label{zphot}

Photometric redshifts are derived using the method described by
\cite{ru01,ru03}. The code fits a linear non-negative superposition of
the 8 templates presented in the left panel of Figure~\ref{templates}
to the broadband photometry using $\chi^2$ minimization. This template
set consist of the empirical E, Sbc, Scd and Im templates from
\cite{co80}, the two least reddened starburst templates from
\cite{ki96}, and a 1 Gyr and 10 Myr \cite{bc03} single stellar
population with a \cite{sa55} initial mass function (IMF). Confidence
intervals on \zp\ are determined using 100 Monte-Carlo simulations, in
which the broadband photometry is varied according to the photometric
uncertainties.

We determine photometric stellar populations properties by fitting
\cite{bc03} stellar population models to the broadband SEDs, with
the redshift fixed to the best-fit \zp. We apply the same grid and
reddening law \citep{ca00} as for the spectroscopic fitting
(\S~\ref{zspec}). The 100 simulations of the photometry in combination
with their best-fit \zp\ are used to derive the confidence intervals
on the stellar population properties (see procedure as described in
\S~\ref{zspec}).

The errors on the photometric redshifts and rest-frame luminosities are
derived independently using the method by
\cite{ru01,ru03}. Due to the independent fitting procedures, we do not
have a combined $\chi^2$ distribution for $M$ and $L_V$. For the
confidence intervals on $M/L_V$ we take the minimum and maximum
$M/L_V$ corresponding to the 68\% best stellar population fits, i.e.,
with the lowest $\chi^2$. All photometric redshifts, stellar population
properties and corresponding confidence intervals are listed in
Table~\ref{tab_phot}.

\subsection{Direct Comparison}\label{dir_comp}

In Figure~\ref{z_comp} we compare \zs\ and \zp\ for all individual
galaxies. We find a scatter of \dzs$=0.13$ and a systematic offset of
\dzs$=0.08$, such that our photometric redshifts are on average too
high. These values are 0.08 and 0.03 respectively for galaxies in the
MUSYC deep fields only (see Table~\ref{tab_dz}). The distributions of
the spectroscopic and photometric redshifts for the full and the deep
sample, are shown in Figure~\ref{z_hist}. Using optical spectroscopy
and the same photometric redshift code and templates, \cite{wu07b}
find a scatter of \dzs$\sim0.06$ and a median offset of \dzs$\sim0.02$
for galaxies with deep NIR photometry in the same redshift range. This
result is similar to ours when comparing only to the galaxies with
deep NIR photometry.

\figk % Figure 10: Histograms N(z) and P(z)
\figl % Figure 11: Broadband photometry and best fits for failures
\figm % Figure 12: Systematics in dz/(1+z)

The broadband photometric redshifts of galaxies with emission lines
have smaller errors than those of galaxies without emission lines,
especially when we exclude the galaxies with shallow NIR
photometry. However, our sample is small, and studies of larger
spectroscopic samples are needed to verify this result. If this result
still holds for larger samples, it probably implies that the quality
of photometric redshifts, which are calibrated using just
emission-line redshifts, may be overestimated.

The broadband photometric redshifts of galaxies with emission lines
have smaller errors than those of galaxies without emission lines (see
also Table~\ref{tab_dz}), especially if we exclude the galaxies with
shallow NIR photometry. However, our sample is small, and studies of
larger spectroscopic samples are needed to verify this result. If this
result still holds for larger samples, it probably implies that the
quality of photometric redshifts, which are calibrated using just
emission-line redshifts, may be overestimated. We stress, however,
that galaxies with emission-line and continuum redshifts do not
precisely present two distinct classes within this sample, as for
certain redshift intervals emission lines are expected in between
atmospheric windows.

Whereas spectroscopic redshifts for blue galaxies can be obtained by
optical spectroscopy as well, for most red, optically faint galaxies
NIR spectroscopy offers the only possibility to obtain a \zs. Previous
studies, our photometric redshift estimates among them, were mainly
calibrated using optical spectroscopy. Therefore, it is interesting to
know whether red and blue high-redshift galaxies show similar scatter
and systematics.  We divide the full sample into two classes using the
DRG criterion \citep[$J-K>2.3$,][]{fr03}. In Figure~\ref{z_comp} we
indicate the DRGs and non-DRGs by red and blue symbols
respectively. We find no significant difference in the scatter or
systematic offset in \dzs\ for blue and red galaxies for the total
sample. When only considering the galaxies with deep NIR photometry,
DRGs seem to have more accurate photometric redshifts than
non-DRGs. We note, however, that due to the redshift dependence of the
$J-K$ color (see Fig.~\ref{rep}), the trend between \dzs\ and \zs\
(which is a logical result of our \zp\ selection) complicates the
interpretation of this test. In the next section we will use
rest-frame colors in order to avoid this effect.

\subsection{Systematics and Catastrophic Failures} \label{sect_sys}

In this section we will examine the different causes for systematics
and catastrophic failures in \dzs\ as identified in the previous
subsection. We define failures as galaxies for which \dzs$>0.1$. In
Figure~\ref{phot} we show the broadband photometry and best-fit
stellar population models, in- or excluding the spectral information,
for all failures. However, for several galaxies the photometric
redshift is poorly constrained -- for example due to degeneracies --
and despite the large discrepancy, \zs\ agrees reasonably well with
the confidence interval of \zp. Thus we discriminate between failures
for which \zs\ is or is not consistent within 3~$\sigma$ with \zp, and
define these as ``good failures'' and ``bad failures''
respectively. This is illustrated in Figure~\ref{sys}a, in which we
show \dzs\ versus the difference between \zp\ and \zs\ in
$\sigma$. For one galaxy with
\dzs$<0.1$ (1256-519), \zp\ and \zs are not consistent within
3~$\sigma$.

Figure~\ref{sys}b shows that all bad failures have shallow NIR
photometry, with an average S/N in $H$ and $K$ of $<10$. This
illustrates that deep NIR photometry is needed to get meaningful
photometric redshifts and corresponding confidence intervals. Not all
galaxies with shallow NIR photometry are bad failures, or even
failures at all, but the probability that a galaxy with shallow NIR
data has a \zp\ which is inconsistent with the spectroscopic redshift
is substantially larger than for galaxies with deep NIR data.

As good failures have similar \dzs\ as bad failures, their confidence
intervals are by definition larger. Large confidence intervals may be
caused by shallow NIR data or model degeneracies. Several good
failures indeed have shallow NIR photometry, and their photometric
redshifts are -- possibly by coincidence -- consistent with their
spectroscopic redshifts. The deviating photometric redshifts of high
S/N failures must have a different origin. In the following we examine
several potential causes for good failures:\smallskip

{\em Deviating NIR colors:} In Figure~\ref{sys}c we relate \dz\ with
the difference between the photometric and spectroscopic $J-H$ color
(see \S\ref{nir_spectra}). As the optical break falls between the $J$-
and $H$-band for $z=2-3$ galaxies, a deviating $J-H$ color may affect
the optical break identification, and thus may place a galaxy at the
wrong redshift. This may have happened for 1030-609 and 1030-2927 (see
Fig.~\ref{spec}). We note that a few galaxies are missing from
Figure~\ref{sys}c as their spectroscopic NIR colors could not be
determined (see Fig.~\ref{q_check}).\smallskip

{\em Emission-line contamination:} Contamination by emission lines
generally reddens the rest-frame optical colors, which may result in a
wrong break identification. Figure~\ref{sys}d shows that for several
galaxies the photometric NIR colors appear redder due to the flux
contribution by emission lines. However, there is only one galaxy
(CDFS-6026, see Fig.~\ref{spec}) with deep NIR photometry, for which
the emission line contamination may be the main cause for a wrong
\zp. \smallskip

{\em K band magnitude:} Due to the shape of the luminosity function,
the fraction of low to high redshift galaxies is much larger at
brighter $K$ magnitudes. Thus, it is expected that the interloper
fraction is larger at the bright end. In Figure~\ref{sys}e we indeed
see that the fraction of good failures compared to the total sample
increases at brighter $K$ magnitudes. This effect can be reduced by
using a luminosity prior in the \zp\ fitting code, to break the
degeneracy in favor of the correct low-redshift solution.\smallskip

{\em Correlations with SED types:} In Figures~\ref{sys}f, g and h we
examine correlations with SED type by comparing \dz\ with rest-frame
$U-V$ color, best-fit age and dust content (all derived from the
spectroscopic information). We use rest-frame $U-V$ colors, instead of
observed $J-K$ colors, as the latter are dependent on redshift (see
\S~\ref{dir_comp}). We find no obvious correlation between \dz\ and
rest-frame $U-V$ color. Nonetheless, Figures~\ref{sys}g and h suggest
a relation with stellar population properties, such that young and/or
dusty galaxies have a higher probability of having a wrong \zp. These
galaxies are in our photometric redshift code generally best-fit by
the 1 Gyr dust-free template (Fig.~\ref{templates}), which places them
at the wrong redshift (see also \S~\ref{sect_prop}). One of these
galaxies is 1256-519, for which \zp\ and \zs are not consistent within
3~$\sigma$ (although \dzs$<$0.1). This may suggest that appropriate
templates (i.e., dusty ones) are missing in our \zp\ procedure.  We
will examine this in more detail in \S~\ref{fix}. \smallskip

All discussed potential causes generally overestimate photometric
redshifts, and place the galaxy at too high redshift. Altogether, this
introduces a systematic effect. In the next section we will examine
the influence of the different causes, by attempting to correct for them.

\subsection{An Empirically-Motivated Template Set}\label{fix}

In order to test the different causes found for the systematics and
failures in \dzs, we refit the photometry attempting to correct for
these effects. First we construct a new template set from our GNIRS
sample. By eye we pick six best-fit SEDs from Figure~\ref{spec}, to be
representative for the $z\sim2.5$ massive galaxy population. These new
templates are presented in the right panel of
Figure~\ref{templates}. We refit the broadband photometry of all
galaxies only allowing these 6 templates, without superposition and
luminosity prior, in steps ($\Delta z$) of 0.02.

In Figures~\ref{newz}a and b we show the comparison between the old
and new \zp, respectively, versus \zs. The galaxies with high S/N NIR
photometry are presented by filled symbols. The new templates reduce
both the systematics and scatter. Remarkably, the improvement is
largest for galaxies with low S/N photometry.

\fign % Figure 13: z_phot vs z_spec for new templates

Although, the newly defined template set yields more accurate
photometric redshifts than the original template set, it provides no
general solution. It is constructed to work especially for massive
galaxies at $2<z<3$, and thus they may not be appropriate when applied
to lower redshift or less massive galaxies. Nevertheless, this test
illustrates the importance of using well-calibrated templates. Similar
template mismatches may also apply to other galaxy populations, and
further exploration is needed in order to better calibrate
high-redshift studies.

In order to test whether errors in the NIR broadband photometry may be
responsible for systematics or may lead to an increase in the scatter,
we attempt to correct the photometry using the NIR spectra. In
Figure~\ref{newz}c we replace the $J$, $H$ and $K$ by the
spectroscopic colors and repeat the original fitting procedure, using
the original templates. The spectroscopic colors do not lead to an
improvement, and in fact even lead to more pronounced scatter and
systematics, in particular for the galaxies with deep NIR
photometry. This might not be surprising, as for these galaxies the
S/N of the $J-H$ photometric colors is generally higher than the S/N
of the spectroscopic colors.

Furthermore, we remove the emission line fluxes from the broadband
photometry before repeating the original fitting
procedure. Figure~\ref{newz}d shows that this does not reduce the
scatter or the systematics.  This may be because several of the
original templates contained emission lines. Also, the massive
galaxies studied in this work all have comparatively weak emission
lines, and contamination may play a much larger role for less massive
galaxies.

\figo % Figure 14: Comparison properties
\addtocounter{figure}{-1}
\figp % Figure 14: Comparison properties

\section{IMPLICATIONS}\label{imp}

The spectroscopic redshifts and higher resolution spectral shapes as
obtained by the NIR spectra provide better constraints on the stellar
population properties. This allows us to test the significance and
reliability of the photometrically derived properties, and examine the
completeness and systematics of photometric samples. In
\S~\ref{sect_prop} we quantify the improved constraints on the stellar
population properties, and identify the driving force behind this
improvement. In \S~\ref{phot_samples} we test photometric samples, and
finally in \S~\ref{prev_studies} we assess previous studies based on
photometric samples.

\subsection{Implications for Derived Properties}\label{sect_prop}

In this section we compare the photometric and spectroscopic stellar
population properties, as listed in Tables~\ref{tab_spec} and
\ref{tab_phot}. Note that the photometric properties are derived using
the original photometric redshifts (\S~\ref{zphot}). Figure~\ref{prop}
illustrates the effect on the properties age, $A_V$, $\tau$, stellar
mass and SFR when including the spectroscopic information. In the left
panels we show the comparison for individual galaxies. The panels in
the second column present the distributions of the photometric and
spectroscopic properties. In the third column we show the distribution
of the difference between photometric and spectroscopic
properties. Finally, in the panels in the fourth column we show the
ratio of the photometric and spectroscopic 1$\sigma$ confidence
intervals.

The differences between the best-fit photometric and spectroscopic
properties in the left diagrams of Figure~\ref{prop} are
striking. $A_V$ has a scatter of $\sim$1 mag, and age, $\tau$ and SFR
each have a scatter of a factor of $\sim$3-4, between the photometric
and spectroscopic values.  Despite the large scatter, for a high
fraction (75-89\%) of the galaxies the photometric properties are
consistent with the spectroscopic properties within 1\,$\sigma$. This
demonstrates the reliability of the large confidence intervals in
Table~\ref{tab_phot}, and supports that the best-fit photometric
properties age, $A_V$, $\tau$ and SFR are poorly determined for
individual galaxies. The spectroscopic properties are better
constrained, but the errors are still substantial, with a factor of
$\sim$2 for age, tau, and SFR and $\sim$0.4 mag for $A_V$. Thus, it is
not surprising that the stellar population properties show a large
scatter, even when the photometric redshift is close to the
spectroscopic redshift.

Although the photometrically and spectroscopically derived best-fit
values are consistent, the photometric and spectroscopic distributions
show systematic differences (Fig.~\ref{prop}, 2nd column). In
\S~\ref{sect_sys} we identified possible correlations between \dzs\
and stellar population properties. Due to the lack of appropriate
templates, the stellar population properties seem to ``converge'' to
the properties of the template that provides the best fit during the
\zp\ procedure. As many galaxies are best-fit by the dust-free 1 Gyr
template, we see a peak at $A_V=0$ and an age of 1 Gyr for the
photometry. The $A_V=0$ peak completely disappears, and the age peak
shifts to lower ages when including spectroscopic information. This
suggest that the distribution of best-fit photometric stellar
population properties, as presented in the 2nd column of
Figure~\ref{prop} may be strongly biased.

\figq %Figure 15: causes for improvement

While it is clear that spectroscopic data significantly improve the
constraints on the stellar population properties, it is interesting to
know whether the improvements are primarily due to the precise
redshift determination or whether they are also caused by the
spectroscopic measurement of the continuum shape. We test the impact
of both effects by modeling the broadband photometry by stellar
population models (see procedure \S~\ref{zspec}), fixing the redshift
to \zs. In Figure~\ref{what} we disentangle the two effects, by
comparing on the horizontal axis the improvement of the stellar
population properties due to the higher resolution spectral shape
(i.e., new modeling results versus spectroscopic properties) and on
the vertical axis the effects of the better constrained redshift
(i.e., photometric properties versus new modeling results). The
corresponding scatter and systematics are listed in
Table~\ref{tab_scatter}. Remarkably, the results for age and $\tau$
are more affected by the higher resolution spectral shape, rather than
by the better constrained redshift. For SFR and $A_V$ both effects are
equally important.

Photometric stellar mass is significantly better constrained, with a
scatter of $\sim2$. Nevertheless, this property is generally
overestimated by a factor of $\sim1.4$. Even if we exclude the
galaxies with low S/N NIR photometry, we still obtain photometric
stellar masses which are on average a factor of $\sim1.3$ too
high. This overestimation is lower than the factor of 2 found by
\cite{sh05} for blue galaxies. However, the values are not directly
comparable, as their mass ratio is derived by using spectroscopic or
photometric redshifts when fitting $UGRK$ and IRAC photometry. In
Figure~\ref{what} we find that the improved constraints on stellar
mass is evenly driven by a better constrained redshift and the higher
resolution spectral shape (see also Table~\ref{tab_scatter}). Thus,
spectroscopic redshifts combined with broadband photometry may still
overestimate the stellar mass, although by a small factor of
$\sim1.1$.

\tabe

In Figure~\ref{prop} we also show the comparison between the
photometrically and spectroscopically derived $M/L_V$, rest-frame
magnitude $V$ and rest-frame $U-V$ color. Of all photometric properties
$M/L_V$ is the best determined, and shows no systematic
offset. Rest-frame $U-V$ colors are in general underestimated by
$\sim0.1$ mag, and show a scatter of $\sim0.2$ for the galaxies with
deep NIR photometry. Figure~\ref{what} shows that the systematics and
scatter are mainly due to the lack of a spectroscopic redshift. This
effect is even worse for rest-frame $V$ magnitudes. As expected,
rest-frame $V$ magnitudes can very accurately be determined once the
redshift is known. Thus, for photometric redshifts this quantity is
less well constrained with an offset of $\sim0.2$ and a scatter of
$\sim0.4$ for the deep NIR photometry.

\figr %Figure 16: implications for photometric studies

Better constrained properties and confidence intervals can also be
obtained by extending the broadband SEDs by mid-infrared Spitzer/IRAC
photometry \citep[e.g.,][]{la05,wu07a}. \cite{wu07a} show that the
distribution of stellar population properties does not change
significantly when IRAC data are added to their {\em UBVIJHK}
photometry. However, for individual galaxies the addition of IRAC can
improve the constraints on the stellar populations significantly. As
we do not have photometric catalogs including the IRAC bands for the
GNIRS galaxies yet, it is difficult to examine whether the IRAC bands
have a similar effect as the higher resolution spectral shape on the
stellar population properties.

Finally, we stress that the derived photometric and spectroscopic
properties are not completely independent, as we use the same optical
photometry for both, and the NIR spectra are calibrated using the NIR
photometry. Furthermore, the confidence levels do not include all
uncertainties, as all photometrically and spectroscopically derived
stellar population properties are based on the same assumptions, among
which a \cite{sa55} IMF, solar metallicity, an exponentially declining
SFH, and the \cite{bc03} stellar population models.

\subsection{Implications for Photometric Samples}\label{phot_samples}

In the previous section we examined the improvement of stellar
population properties when including the spectroscopic
information. However, due to the different redshift distributions, the
photometric and spectroscopic properties as presented in
Figure~\ref{prop} should not be compared directly to examine the
implications for galaxy samples. Due to degeneracies between redshift
and stellar population properties, the systematic effects may even be
worse for complete samples, than for individual galaxies. Thus, it is
important to consider the same redshift interval in order to examine
the systematics of photometric samples.

In the top panels of Figure~\ref{impl} we show the photometric and
spectroscopic properties stellar mass, rest-frame $U-V$ color, and
observed $J-K$ and $R-K$ color versus redshift of the total GNIRS
sample. The photometrically derived properties are presented by open
diamonds and the filled squares represent the spectroscopically
derived properties. Hence, all galaxies are shown twice. This figure
illustrates how photometric and spectroscopic properties of galaxies
in a certain redshift interval differ, and which galaxies fall out and
in the targeted redshift interval ($2.0<z<2.7$) when spectroscopic
information is included. For example, several of the galaxies for
which the photometric redshift and hence stellar mass were
substantially overestimated, move to $z<2$, and thus will not have a
large effect on the properties of the $2.0<z<2.7$ sample. We note,
however, that our spectroscopic sample is not complete, as galaxies
with $z_{\rm phot}<2.0$ and $z_{\rm phot}>2.7$ may scatter into the
spectroscopic sample. Although the photometric sample contains 10
galaxies with $z_{\rm phot}>2.7$, these galaxies may not be
representative for all galaxies with $z_{\rm phot}>2.7$ and
$2.0<z_{\rm spec}<2.7$.

In the bottom panels in Figure~\ref{impl} we compare the photometric
and spectroscopic distributions of stellar mass, $J-K$, $R-K$ and
rest-frame $U-V$ of the $2.0<z<2.7$ galaxy sample. The first panel
illustrates that photometric stellar masses in general are slightly
overestimated. In order to compare how the properties of a
mass-limited sample would change, we use a mass-cut of $>10^{11}
M_{\odot}$ ($M_{\rm spec}$ for the ``spectroscopic'' sample and
$M_{\rm phot}$ for the ``photometric'' sample) in addition to the
redshift interval for the remaining properties. The second panel
demonstrates that spectroscopically derived rest-frame $U-V$ colors of
a mass-limited sample at $2.0<z<2.7$ are slightly redder ($\sim$0.1
mag) than inferred from just the photometry. Furthermore, the
distribution is peaked around $U-V\sim0.75$, possibly indicating a red
sequence (M. Kriek et al. 2007 in preparation). Also, the median $J-K$
color as inferred from the photometry is $\sim0.1$~mag too blue, while
the median $R-K$ is $\sim0.15$~mag too red. This may reflect the fact
that dusty galaxies (see \S~\ref{sect_sys}) with relatively red $R-K$
and blue $J-K$ colors fall out of the sample, and move to lower
redshift. Nonetheless, the photometric and spectroscopic distributions
are remarkably similar for both $J-K$ and $R-K$.

The systematics in both the photometric mass and redshift distribution
introduce a systematic in number of massive galaxies in a certain
redshift interval, as shown in Figure~\ref{impl}. As both the
photometric masses and redshifts at $z\sim2.5$ are generally
overestimated, the number of massive galaxies and the mass density
will be lower than inferred from the photometry. Our sample is too
small to directly quantify this effect. However, we can use the
originally photometric sample to estimate the change of the number of
massive galaxies in certain redshift bins.

We construct a $K$-selected ($K<21.3$) photometric galaxy sample from
the deep MUSYC catalog, and correct this sample using the coupled
distributions of \dz\ and $M_{\rm spec}/M_{\rm phot}$. For these
distributions we solely use the GNIRS galaxies in the deep fields
(excluding ECDFS galaxies). As galaxies will scatter in and out of the
relevant redshift interval ($2<z<3$) from both low and high redshift,
we start with a photometric sample over a larger redshift interval
($1.5<z<4.0$). We randomly apply the coupled distributions to the
original sample, and repeat this process 1000 times. From the original
sample and all 1000 simulations we construct a mass-limited sample
($>10^{11} M_{\odot}$). We apply the mass-cut after correcting, as
galaxies that do not make the cut in the original sample can still
scatter into the corrected sample. The depth of the $K$-selected
catalog allows extraction of an almost complete mass-limited sample
\citep{vd06}. Finally, we divide the sample in redshift bins of
$\Delta z=$0.1 and average the simulation to construct the final
corrected number distribution. Figure~\ref{model_n} presents both the
original and corrected number distributions of a mass-limited sample
as a function of redshift.

The difference in the original and corrected distribution in
Figure~\ref{model_n} shows that photometric studies may overestimate
the number of massive galaxies by a factor of $\sim$1.3 at
$2<z<3$. However, this test is a simplification. First, we used the
same coupled distributions of \dz\ and $M_{\rm spec}/M_{\rm phot}$
randomly for all galaxies, while we know that there are correlations
between \dzs\ and S/N of the photometry, $K$ magnitude, SED type
etc. (see \S~\ref{sect_sys}). Our spectroscopic sample is too
small to quantify these correlations in an applicable form. Second,
our sample and thus the distributions of \dzs\ and $M_{\rm
spec}/M_{\rm phot}$ may be incomplete. For example, it may be that we
miss galaxies with a \zp$\sim4$ and a \zs$\sim2$. A larger sample over
a larger redshift range in needed to better understand the
systematics.

\subsection{Implications for Previous Studies}\label{prev_studies}

At this moment most studies of massive galaxies beyond $z>1.5$ rely on
photometric redshifts. Thus, the systematics found in this work will
have consequences for previous studies, especially to those that use
the same or a similar method to derive photometric redshifts. In this
section we will discuss several studies and how our findings may affect
their results.

Using a photometric mass-limited sample ($M>10^{11} M_{\odot}$)
\cite{vd06} recently showed that massive galaxies at $2<z<3$ have red
rest-frame optical colors, and found a median $J-K$ and rest-frame $U-V$
color of 2.48 and 0.62 respectively. This study also contained the
deep MUSYC fields (in addition to several other deep fields) and used
the same photometric redshift code as used in this paper. In the
previous section we showed that the distributions of $J-K$ and $R-K$
barely changes when including the spectroscopic information. The
rest-frame $U-V$ color is slightly redder and more peaked, but this
difference is consistent with the photometric errors. Thus, our
spectroscopic study does not change, but supports the overall
conclusion by \cite{vd06} that the high mass end at $2<z<3$ is
dominated by red galaxies.

\figs %Figure 17: number of massive galaxies

Although the colors of massive galaxies at $2<z<3$ barely change when
including spectroscopic information, the number of massive galaxies
decreases (see \S~\ref{phot_samples}). This will be of importance for
our understanding of the evolution of the mass density
\citep[e.g.,][]{di03,ru03,ru06,dr05}. Due to the numerous low redshift
spectroscopic surveys, photometric redshifts below $z=1$ are well
calibrated and have small uncertainties in the order of
\dzs$\sim0.03$. This results in well-calibrated masses. However, as
shown in this paper, photometric redshifts beyond $z>1.5$ are
generally poorly calibrated and may show strong systematics.  These
systematics may result in too high redshifts, and thus too high
stellar masses. For the photometric redshift code and template set
used in this paper, the evolution of the global stellar mass density
between $z\sim2.5$ and $z\sim0$ would have been underestimated by a
factor of $\sim$1.3, due to the combination of redshift and mass
systematics. However, this factor is almost entirely driven by
overestimates of photometric redshifts. As other studies may use
different
\zp\ codes and templates sets, they may not suffer from this effect.

Our results will also affect galaxy clustering studies
\citep[e.g.,][]{da03,qu07a,fo07}. Galaxy clustering at high redshift is
measured using the angular two-point correlation function. The angular
correlation function can be converted into the spatial correlation
function using the Limber equation. The main ingredient in this
equation is the redshift distribution. \cite{qu07a} use the summed
redshift probability distribution, in order to account for different
\zp\ errors of the individual galaxies. In the two bottom panels of
Figure~\ref{z_hist} we compare the summed probability distributions of
the photometric redshifts of the GNIRS galaxies for both the full and
deep sample, with the spectroscopic distribution. For the continuum
redshifts we use the probability distribution as well. Although the
median of the distribution for both samples is slightly offset in
redshift, the width of the distribution is similar. Because the Limber
deprojection is more dependent on changes to the width of the
distribution than the central value, the clustering results will not
change significantly when using spectroscopic redshifts. One can argue
that this might be coincidence and further study is necessary to
confirm this.

\section{SUMMARY}\label{sum}

In this paper we present our complete NIR spectroscopic survey for
$K$-selected galaxies at $z\sim2.3$. We acquired high quality NIR
spectra (1-2.5 $\mu$m) with GNIRS for a total sample of 36
galaxies. The galaxies were selected to be bright in $K$ ($<$19.7) and
have photometric redshifts in the range $2.0\lesssim z\lesssim
3.0$. All galaxies have photometric stellar masses $>10^{11}
M_{\odot}$. The distribution of observed $J-K$, $R-K$ and photometric
rest-frame $U-V$ colors are similar as those of a photometric
mass-limited sample extracted from the deep MUSYC fields. This
suggests that our spectroscopic sample is representative for a
mass-limited ($>10^{11} M_{\odot}$) sample at $2<z_{\rm phot}<3$. Our
sample is not representative for a $K$-bright galaxy sample, as our
redshift distribution is slightly higher. Nevertheless, the
distribution of photometric rest-frame $U-V$ colors is similar for the
GNIRS sample and the full $K$-bright photometric sample.

We successfully derived spectroscopic redshifts for all galaxies. For
19 of the galaxies we detected rest-frame optical emission lines, which
provided us with accurate redshift measurements. For the remaining
galaxies we derived redshifts by modeling the continuum spectra in
combination with the optical photometry. We tested this method using
the emission-line galaxies, and found a scatter of $\Delta
z/(1+z)=0.019$, and no significant systematic offset. The continuum
redshifts are more accurate for optically red galaxies, with large
Balmer/4000 \AA\ breaks, and for redshifts for which the break falls
in an atmospheric window.

Comparison with spectroscopic redshifts shows that the original
photometric redshifts are generally overestimated by $\Delta
z/(1+z)=0.08$, and have a scatter of $\Delta z/(1+z)=0.13$. Both the
systematic and the scatter are worse for galaxies with shallow NIR
photometry ($K\sim20$). For the galaxies with deeper NIR photometry
($K\sim21$) we find a systematic offset of $\Delta z/(1+z)=0.03$ and a
scatter of $\Delta z/(1+z)=0.08$.

In addition to shallow NIR photometry, the lack of well calibrated
galaxy templates is another important cause for the large systematics
and scatter of photometric redshifts. Especially, dusty and/or young
galaxies have high \dzs. This may not be surprising as the original
template set of our photometric redshift code lacked these
galaxies. We almost completely remove the systematic offset when
repeating the fitting procedure replacing the original templates by 6
SEDs chosen from fits to the spectroscopic sample. We examine other
possible causes for the scatter and systematics in \dzs\ as
well. Emission line contamination is only of minor importance for this
sample and may affect the \zp\ for some individual
galaxies. Furthermore, we find a correlation with $K$-band magnitude,
such that $K$-bright galaxies in general have larger \dzs\ than those
faint in $K$. This effect may be diminished by using a luminosity
prior in the photometric redshift code. Although improved photometric
redshift codes and better templates may remove the systematics and
decrease the scatter, photometric redshifts remain limited by the low
resolution spectral shape, and are not likely to surpass a scatter of
\dzs$=0.0$5 at the targeted redshift range -- even for the galaxies
with deep NIR photometry. We note that obtaining photometry in
narrower bands and for a larger number of filters may lead to better
results than has typically been the case to date.

The spectroscopic stellar population properties, as derived by
modeling the spectral continuum shape together with optical
photometry, allow us to examine the significance and reliability of
the stellar population properties derived from the photometry
alone. Strikingly, the properties age, SFR, $\tau$ show a scatter of a
factor of$\sim3-4$ between the photometrically and spectroscopically
derived properties. For $A_V$ this scatter is $\sim1$ mag. The
improvement in age and $\tau$ is driven primarily by the higher
resolution spectral shape, and less so by the better constrained
redshift. For SFR and $A_V$ both effects are about equally
important. The uncertainties in the photometric modeling seem
reasonable, as for $\sim3/4$ of the galaxies the photometric and
spectroscopic properties are consistent within
1$\sigma$. Nevertheless, the photometric and spectroscopic
distributions are quite different, accounting for their errors. This
may be caused by the fact that age, $A_V$ and $\tau$ suffer from
systematics, due to degeneracies with the photometric redshift. For
example, the photometric stellar population properties for several
galaxies converge to the properties of best-fit templates, used in the
photometric redshift code. This all implies that the best-fit
photometric properties and especially their distribution are strongly
biased.

The photometrically derived stellar mass and mass-to-light ($M/L_V$)
ratio are best determined and show a scatter of a factor of $\sim2$
with their spectroscopic analogues. However, the systematic offset in
\dzs\ results in a systematic offset in $M_{\rm phot}/M_{\rm spec}$,
such that $M_{\rm phot}$ is about 30\% too high for the galaxies with
deep NIR photometry. This systematic is canceled out in $M/L_V$, as
the $L_{V,\rm phot}$ suffers from similar systematics as $M_{\rm
  phot}$. The improvement for both stellar mass and
$M/L_V$ is about equally driven by the higher resolution spectral
shape and the better constrained redshift.

Rest-frame $V$ magnitudes and $U-V$ colors are the derived properties
which are most sensitive to photometric redshift uncertainties and low
S/N photometry.  The scatter between the photometrically and
spectroscopically derived properties is 0.4 and 0.2 mag for $V$ and
$U-V$ respectively for galaxies with deep NIR photometry, and both
increase by 50\% when we include the galaxies with shallow
photometry. The systematics in the photometric redshifts cause a
systematic offset of -0.2 and -0.1 mag for $V$ and $U-V$ for galaxies
with deep NIR photometry. Thus, for both the scatter and the
systematics, the improvement is almost completely driven by better
constrained redshifts.

In order to assess the impact of the identified systematics on
complete photometric samples, we compared photometric and
spectroscopic samples in the same mass and redshift
range. Surprisingly, we found that the distribution of $J-K$ and $R-K$
color barely changes for massive galaxies ($>10^{11}M_{\odot}$) at
$2.0<z<2.7$. Thus, the spectroscopy supports previous studies that red
galaxies dominate the high mass end at $z\sim2.5$. However, the
combination of photometric redshift and stellar mass systematics also
affect the number of massive galaxies. The photometric redshift code
and template set used in this work overestimates the number of massive
galaxies by factor of $\sim$1.3 at $2<z<3$, resulting in an
underestimation of the stellar mass density between $z\sim2$ and
$z\sim0$.

Overall, our spectroscopic survey demonstrates the large uncertainties
of best-fit photometric properties, and the necessity for accurate
calibration of photometric studies. Although this study is a step
forward in understanding the systematics in photometric studies of
massive galaxies at $z>2$, larger spectroscopic samples over a larger
redshift range are needed to fully map the systematics and accurately
calibrate photometric studies. Furthermore, this study only applies to
the high-mass end of the high-redshift galaxy population. Less massive
galaxies may have other systematics, although that remains to be
explored.

\acknowledgments We thank Gemini Observatory for their enthusiasm and
support of our program. We also thank the anonymous referee for
valuable comments on the manuscript. This research was supported by
grants from the Netherlands Foundation for Research (NWO), the Leids
Kerkhoven-Bosscha Fonds, National Science Foundation grant NSF CAREER
AST-044967, and NASA LTSA NNG04GE12G. DM is supported by NASA LTSA
NNG04GE12G. EG is supported by an NSF Astronomy \& Astrophysics
Postdoctoral Fellowships under award AST-0401547. PL acknowledges
support from Fondecyt Grant no. 1040719.
  
%\newpage

%\newpage

%\taba
%\tabb
%\tabc
%\tabd
%\tabe

%\newpage
%\clearpage

%\figa
%\figb
%\figc
%\figd
%\fige
%\addtocounter{figure}{-1}
%\figf
%\figg
%\figh
%\figi
%\figj
%\figk
%\figl
%\figm
%\clearpage
%\fign
%\figo
%\addtocounter{figure}{-1}
%\figp
%\figq
%\figr
%\figs

\end{document}